\documentclass[pra,aps,twocolumn,preprintnumbers,showpacs]{revtex4}
\usepackage{latexsym,epsfig,amssymb,amsfonts,amsmath,graphicx,bbm,mathptm}
\usepackage[utf8]{inputenc}
\usepackage{hyperref}
\usepackage{enumitem}

\hypersetup{
	colorlinks   = true
}

\newcommand{\ket}[1]{\left | \, #1 \right \rangle}

\newcommand{\bra}[1]{\left \langle #1 \, \right |}

\newcommand{\av}[1]{\langle #1\rangle}

\newcommand{\vac}{\ket{\textrm{vac}}}
\newcommand{\comm}[2]{\left[ #1 , #2  \right]}

\newcommand{\eqr}[1]{Eq.~(\ref{#1})}
\newcommand{\fir}[1]{Fig.~\ref{#1}}
\newcommand{\secr}[1]{Sec.~(\ref{#1})}

\newcommand{\mi}{\mathrm{i}} 
\newcommand{\di}{i}          

\makeatletter
\newsavebox{\@brx}
\newcommand{\llangle}[1][]{\savebox{\@brx}{\(\m@th{#1\langle}\)}%
  \mathopen{\copy\@brx\kern-0.5\wd\@brx\usebox{\@brx}}}
\newcommand{\rrangle}[1][]{\savebox{\@brx}{\(\m@th{#1\rangle}\)}%
  \mathclose{\copy\@brx\kern-0.5\wd\@brx\usebox{\@brx}}}
\makeatother

\allowdisplaybreaks

\begin{document}

\title{Controllable Finite-Momenta Dynamical Quasicondensation in the Periodically Driven One-Dimensional Fermi-Hubbard Model}

\author{Matthew W. Cook$^{1}$, and Stephen R. Clark$^{2,3}$}
\address{$^1$Department of Physics, University of Bath, Claverton Down, Bath BA2 7AY, U.K.}
\address{$^2$H. H. Wills Physics Laboratory, University of Bristol, Bristol BS8 1TL, UK.}
\address{$^3$Max Planck Institute for the Structure and Dynamics of Matter, University of Hamburg CFEL, Hamburg, Germany.}

\date{\today}

\begin{abstract}
In the strongly interacting limit of the Hubbard model localized double-occupancies form effective hard-core bosonic excitations, called a doublons, which are long-lived due to energy conservation. Using time-dependent density-matrix renormalisation group we investigate numerically the dynamics of doublons arising from the sudden expansion of a spatially confined band-insulating state in one spatial dimension. By analysing the occupation scaling of the natural orbitals within the many-body state, we show that doublons dynamically quasicondense at the band edges, consistent with the spontaneous emergence of an $\eta$-quasicondensate. Building on this, we study the effect of periodically driving the system during the expansion. Floquet analysis reveals that doublon-hopping and doublon-repulsion are strongly renormalised by the drive, breaking the $\eta-$SU(2) symmetry of the Hubbard model. Numerical simulation of the driven expansion dynamics demonstrate that the momentum in which doublons quasicondense can be controlled by the driving amplitude. These results point to new pathways for engineering non-equilibrium condensates in fermionic cold-atom experiments and are potentially relevant to driven solid-state systems.  
\end{abstract}

\pacs{03.67.Mn, 03.67.Lx}

\maketitle

\section{Introduction}

Strongly correlated quantum systems are well known to exhibit a wide variety of novel phenomena like antiferromagnetism, the fractional quantum Hall effect and high-$T_c$ superconductivity. If such systems are driven out of equilibrium the emerging physics is expected to be richer still. So far only a small portion of this phenomenological landscape has been explored experimentally, and even less is understood theoretically. As a result the non-equilibrium dynamics of quantum many-body systems one of the most challenging branches of modern physics. Yet it is attracting growing attention due to the spectacular experimental advances in numerous complex quantum systems, ranging from cold-atom \cite{Greiner2002,Bloch2008}, photonic \cite{Christodoulides2003,Rechtsman2013,Noh2017}, optomechanical \cite{Ludwig2013} and condensed matter platforms \cite{Fausti2011,Stojchevska2014,Wang2013}. In particular the intense interest stems from the ability to implement controllable strong perturbations to a system and subsequently measure its properties in real-time with a resolution commensurate with the intrinsic microscopic time scales \cite{Aoki2014}. This capability has opened up new spectroscopies for probing non-equilibrium dynamics as well as new approaches for manipulating them \cite{Gianetti2016}. 

One exciting example of this has been the enormous progress over the past decade in ultra-fast THz pump-probe experiments on solid-state systems \cite{Orenstein2012,Nicoletti2016,Giannetti2016}. By strongly driving low-energy structural or electronic degrees of freedom of a solid \cite{Rini2007} the ultra-fast melting of equilibrium long-ranged order, such as charge-density waves \cite{Fausti2011,Schmitt2008,Miyano1997,Cavalleri2001,Perfetti2008}, magnetic order \cite{Stanciu2007,Ehrke2011}, and orbital order \cite{Forst2011a} has been demonstrated. Even more remarkably, recent experiments have also used strong external modulations to {\em induce} superconducting order far from equilibrium in several different materials \cite{Hu2014,Kaiser2014a,Mankowsky2014,Mitrano2016}. The long-term goal of this approach is ultimately to design and control quantum materials properties ``on demand" by using driving to stabilize order that is otherwise inaccessible thermally \cite{Basov2017,Mankowsky2016}. From a theoretical perspective these experiments raise important fundamental questions about what mechanisms exist for the emergence of order in driven systems, some of which have been explored in a number of recent studies \cite{Denny2015,Sentef2016,Knap2016,Subedi2014,Coulthard2017,Kennes2017}. 

Complementary to real materials, there have been equally spectacular experiments with systems of ultra-cold atomic gases in optical lattices \cite{Hofstetter2018,Jordens2008,Schneider2008}. These `synthetic' solids provide near ideal quantum simulations of Hubbard-like Hamiltonians. Consequently they offer unique perspectives on non-equilibrium dynamics of interacting systems, owing to the unprecedented tunability of their time-dependent hopping amplitudes and inter-particle interactions, as well as the ability to engineer novel initial states \cite{Bloch2008}. Cold-atom systems have thus opened up many long-standing non-equilibrium problems to exquisite experimental scrutiny, such as quenching across quantum phase transitions \cite{Greiner2002}, controlling exchange interactions~\cite{Anderlini2007,Trotzky2009}, transport effects with strong interactions \cite{Brantut2012,Stadler2012,Schneider2012,Lebrat2018} as well as the influence of integrability \cite{Kinoshita2006,Langen2017} and many-body localization \cite{Schreiber2015} on closed system thermalization \cite{Eisert2015}. Of particular relevance to our work are recent experiments~\cite{Gorg2018,Messer2018} where the effect of periodic driving on the basic interactions in fermionic cold-atom systems have been unravelled.

Motivated by all these developments, here we focus on a particularly intriguing example of the spontaneous emergence of order, namely {\em dynamical quasicondensation}, predicted \cite{Rigol2004} and observed experimentally in cold-atom systems \cite{Vidmar2015}. Our aim is to assess whether a similar effect occurs with fermions, with broader implications for electronic systems. Dynamical quasicondensation manifests from a unit-filled Mott insulating region of strongly interacting bosons confined to the centre of a one-dimensional optical lattice with spacing $a$. Upon quenching the confinement this inhomogeneous initial state expands out into the surrounding empty lattice. It is found that the momentum distribution of the hard-core bosons quickly develops sharp peaks at quasimomenta $q_{\rm c} = \pm \pi/2a$ and power-law decaying spatial correlations, signalling unconventional current-carrying quasicondensation \cite{Rigol2004,Rigol2005}. This unusual phenomenon has since been explained by showing the time-evolved state in this case is always an eigenstate of a time-dependent emergent Hamiltonian, nontrivially related to the underlying Hamiltonian governing the system \cite{Vidmar2017}. 

In this work we examine whether dynamical quasicondensation occurs with the expansion of a spatially confined band insulating state in the fermionic Hubbard model. In this case the initial state is a cluster of double occupations (doublons), which for strong repulsive interactions is highly energetic. However, when this energy far exceeds the single particle bandwidth energy conservation demands that the decay of doublons occur through multi-particle scattering processes that are exponentially suppressed with increasing energy \cite{Sensarma2010}. The stability of such repulsively bound pairs has been confirmed experimentally \cite{Winkler2006,Strohmaier2010,Sensarma2010} and their distillation dynamics \cite{Heidrich-Meisner2009} has recently been observed \cite{Xia2015,Scherg2018}. Since doublons are long-lived bosonic quasiparticles the question of whether they Bose (quasi)-condense has been addressed. It was found that they do condense at the band-edge $q_{\rm c} = \pm \pi/a$, leading to adiabatic proposals \cite{Rosch2008,Kantian2010} for generating the much sought-after $\eta$-condensate \cite{Yang1989}. Using time-dependent density matrix renormalization group (td-DMRG) methods \cite{Vidal2003,White2004,Daley2004,Alassam2017,Schollwock2011} we demonstrate here that the sudden expansion of the band insulator also undergoes dynamical $\eta$-quasicondensation with definitive signatures emerging within 10's of hopping times.
  
We significantly expand the relevance of quench induced dynamical quasicondensation by examining its interplay with a simultaneously applied strong periodic driving. The effect of time periodic external fields can be captured by Floquet theory \cite{Shirley1965,Eckardt2015,Bukov2015} and has been successfully used to predict and explain wide ranging phenomena in condensed matter and cold-atoms systems. Seminal examples include induced topological effects for cold-atoms via optical lattice modulation \cite{Jotzu2014,Struck2012} or exposure to circularly polarised light \cite{Grushin2014,Kitagawa2011} in solids, dynamical localization induced Mott transitions \cite{Dunlap1986,Eckardt2005}, effective repulsive to attractive interaction conversion by band-flipping \cite{Tsuji2011}, and the renormalization of the super-exchange interaction \cite{Mentink2014,Coulthard2017,Mendoza-Arenas2017}. Here we use Floquet theory to show that the driving breaks the $\eta$-SU(2) symmetry of the Hubbard model \cite{Kitamura2016} and that by tuning above resonance a wide range of doublon dynamics is realizable, including one where they are non-interacting and directly mimic hard-core bosons \cite{Rigol2004}. We compare this effective theory to td-DMRG numerical calculations to show that even finite frequency driving applied on short time scales can accurately control the momentum at which quasicondensation occurs.

This paper is organised as follows. In \secr{sec:model}, we introduce the driven Hubbard model and initial state that is the focus of this work. \secr{sec:effective} is devoted to deriving and analysing a simpler effective theory. This begins in \secr{sec:effective_undriven_model} and \secr{sec:effective_driven_model} by employ a combination of a strong coupling expansion and Floquet theory to reduce the full driven problem into a quench of an effective doublon Hamiltonian with a driving amplitude dependent anisotropy. Building on this \secr{sec:doublon_melting} and \secr{sec:quasicondense_signatures} then describe td-DMRG calculations that solve the quench dynamics of this effective Hamiltonian for large systems. \secr{sec:driven_full_system} then returns to analyse numerically the full driven Hubbard model by confirming in \secr{sec:undriven_hubbard} and \secr{sec:driven_hubbard} that the behaviour seen in the effective model also manifests when the interactions are moderate and when a finite frequency drive is applied abruptly. In \secr{sec:cold_atom_setup} we examine the realisation of the full driven Hubbard model with current cold-atom experiments and analyse how signatures of driving controlled dynamical quasicondensation in the momentum distribution will appear in real-time measurements. Finally we conclude in \secr{sec:conclusion}.

\section{Model and setup} \label{sec:model}
In this work we investigate dynamics of the fermionic Hubbard model in one dimension (1D) with $L$ sites and open boundaries. The Hamiltonian is $H_{\text{hub}} = -t H_{\text{kin}} + U H_{\text{int}}$ composed of single-band kinetic and on-site interacting contributions (taking $\hbar = 1$ throughout)
	\begin{align}
		\begin{split}
			\label{EQN:Hubbard_Model}
		H_{\text{kin}} &= \sum_{i=1}^{L-1}\sum_\sigma\left( c^\dagger_{i \sigma}c_{i+1,\sigma} + \text{h.c.} \right), \\
		H_{\text{int}} &= \sum_{i=1}^L n_{i , \uparrow} n_{i,\downarrow},
		\end{split}
	\end{align}
with the hopping amplitude $t$, and the on-site repulsion $U \geq 0$.	
Here, $c^\dagger_{i,\sigma}$ operators create a spin-$\sigma = \{\uparrow,\downarrow\}$ fermion localized on lattice site $i$, with the corresponding density for spin-$\sigma$ on the site being $n_{i,\sigma} = c^\dagger_{i,\sigma}c_{i,\sigma}$ and $n_i = n_{i,\uparrow} + n_{i,\downarrow}$. 

We will consider the regime of strong interactions $U \gg t$ large enough that strongly correlated effects become readily apparent both in and out of equilibrium \cite{Scherg2018}. We take the initial state of the system to be the ground state of $H_{\text{hub}} + H_{\text{con}}$, where $H_{\text{con}}=\sum_{i=1}^L v_i n_i$ is a box confinement potential, with $v_i = 0$ for sites $i \in O$ inside a contiguous patch of the chain $O$ of size $N= |O|$, and $v_i \gg U$ otherwise. We assume the system has a total of $N_\uparrow = N_\downarrow = N$ electrons making the region $O$ doubly-filled and the ground state 
	\begin{align}
		\ket{\psi_{\text{init}}} = \left(\prod_{i \in O } c^\dagger_{i \uparrow} c^\dagger_{i \downarrow}\right) \ket{\text{vac}},
	\end{align}
which is a spatially confined band insulator, a portion of which is depicted in \fir{FIG:Setup}(a).

We consider the dynamics of this system in real time $\tau$ simultaneously subject to two different time-dependent perturbations which together give a Hamiltonian
	\begin{align} \label{EQN:Driven_Hubbard_Model}
		H_{\text{full}}(\tau) = H_{\text{hub}} + \theta(-\tau)H_{\text{con}} + \theta(\tau)H_{\text{drv}}(\tau).
	\end{align}
The first perturbation describes the sudden switch off at $\tau = 0$ of the confining potential, resulting in a quench whose expansion dynamics melts $\ket{\psi_{\text{init}}}$, as depicted in as depicted in \fir{FIG:Setup}(b). The second perturbation describes the abrupt switch on at $\tau = 0$ of an external periodic driving
	\begin{align} 
	\label{EQN:Ham_Drv_Part}
		H_{\text{drv}}(\tau) =  \frac{A}{2} \cos(\Omega \tau) \sum_i (-1)^i n_{i},
	\end{align}
describing an oscillating on-site energy alternating between sub-lattices with a frequency of $\Omega$ and amplitude $A$, as illustrated in \fir{FIG:Setup}(c). We will be considering high frequency cases where $\Omega \gtrsim U$ is the largest energy scale in the system. The driving term $H_{\text{drv}}(\tau)$ can be realizable in cold-atom experiments by laterally modulating in time the position of a zig-zag optical lattice configuration~\cite{Zhang2015}.

\begin{figure}[t]
\centering
\includegraphics{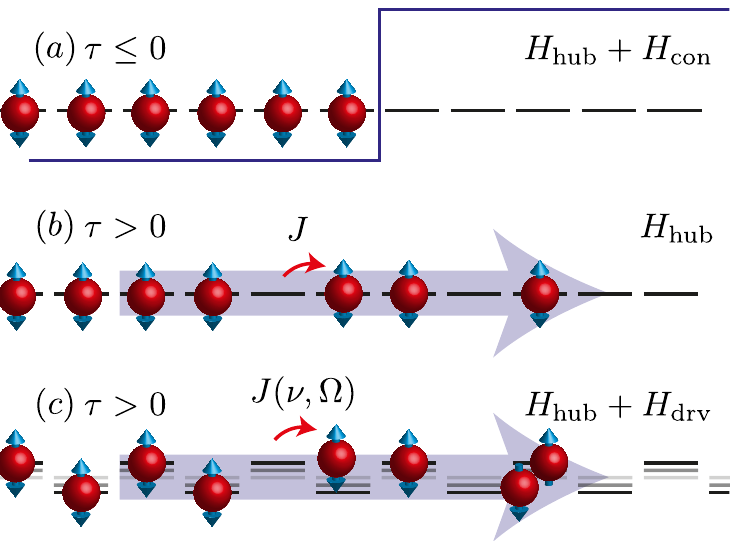}
\caption{Schematic of the setup. (a) The initial state is created with a strong confining potential $H_{\text{con}}$ to produce a band insulator of double occupations. (b) The confining potential is then removed allowing doublons to expand in to the empty region. (c) A driving potential $H_{\text{drv}}$ is switched on simultaneous with the confinement quench. \label{FIG:Setup}}
\end{figure}

We calculate the time-evolution of this system using td-DMRG~\cite{Vidal2003,White2004,Daley2004,Alassam2017,Schollwock2011} with a time-step $\delta\tau < T/50$ where $T = 2\pi/\Omega$ is the drive period, and a matrix product bond dimension $\chi > 1000$ sufficient to ensure the discarded weight $\epsilon_{\rm disc} < 10^{-6}$. In our numerical calculations we take the region $O$ to be the right half of a chain with $L \geq 2N$ giving an initial state of the form $\ket{\psi_{\text{init}}} = \ket{\Updownarrow \Updownarrow \Updownarrow ... \Updownarrow 0 ... 0 0 0 }$. This asymmetric setup, where particles can only expand in one direction, allows access to longer timescales compared to the more experimentally motivated case where $O$ is located at the centre of the system that is overall twice as large. Once $L$ is large enough that no reflections occur at either open boundary in the simulated time we find the same results as those of the symmetric setup \cite{Meisner2008}. 

\section{Quenched effective model} \label{sec:effective}
Accurately computing the dynamics of the full driven Hubbard model $H_{\text{full}}(\tau)$ on long time-scales is extremely challenging. For this reason, and to give a deeper understanding of the physics, we begin our analysis by deriving an equivalent quench problem for a simpler effective model, applicable in the regime of the strong interactions $U \gg t$ and high-frequency $\Omega \gtrsim U$ drives. 

\subsection{Undriven system} \label{sec:effective_undriven_model}
For the Hubbard model $H_{\text{hub}}$ in the strongly interacting limit, doublons are known to be repulsively bound long-lived excitations, owing to their binding energy far exceeding the single-particle bandwidth~\cite{Winkler2006,Strohmaier2010}. The decay of doublons within the Hubbard model, in the presence of background holes, is dominated by multi-particle processes that generate many particle-hole pairs. A diagrammatic perturbative argument~\cite{Sensarma2010} indicates that the rate of doublon decay $\Gamma$ is exponentially suppressed with $U/t$ as
	\begin{align} 
      \Gamma \sim C t\, \exp[-\alpha (U/t) \log(U/t)], 
	\end{align}
where the constants are $\alpha \approx 0.82$ and $C \approx 1.6$. This motivates examining the expansion dynamics of $\ket{\psi_{\text{init}}}$ using an effective model that explicitly conserves doublons. 

Given $\ket{\psi_{\text{init}}}$ contains a maximally localized doublon domain for its filling, it is a highly excited state of $H_{\text{hub}}$ with an energy $E \approx UN$. This places it predominantly in the highest well-isolated band of Hubbard eigenstates when $U \gg t$. An effective Hamiltonian $H_{\text{eff}}$ describing just this highest set of eigenstates is derived in a standard perturbative approach accounting for virtual transitions to lower eigenstates. This is entirely analogous to the well-known derivation of the isotropic antiferromagnetic Heisenberg spin model describing the lowest-lying eigenstates of the half-filled Hubbard model~\cite{Anderson1959}. The result reveals that the system is governed to second order in $t/U$ (up to a constant) by 
	\begin{align}\label{Eqn:UndrivenEffectiveHam}
		H_{\text{eff}} = \mathbbm{P}	 \left[ \frac{J_0}{2}\left(H_{\text{hop}} + H_{\text{rep}}\right)\right] \mathbbm{P},
	\end{align}
where $\mathbbm{P}$ is the projector onto real-space configurations containing no singly occupied sites and
    \begin{align} 
		\begin{split}
		H_{\text{hop}} & =  \sum_{i=1}^{L-1} \left(d^\dagger_i d_{i+1} + {\rm h.c.}\right), \\
		H_{\text{rep}} & =  \sum_{i=1}^{L-1} \Big(n_{d,i}(1-n_{d,i+1})+ {\rm h.c.}\Big).
		\end{split}
	\end{align}
Here the allowable empty $\ket{0}$ or doubly occupied $\ket{\Updownarrow}$ local charge states are equivalent to the absence or presence of a hard-core boson described by the doublon creation operator $d^\dagger_i = c^\dagger_{i \downarrow} c^\dagger_{i \uparrow}$ obeying $(d^\dagger_i)^2 = 0$ and associated number operator $n_{d,i} = d^\dagger_i d_i$. For $U>0$, the doublon hopping term $H_{\text{hop}}$ gives a single-doublon band with its minimum at the zone boundary $q=\pm\pi/a$, while the interaction term $H_{\text{rep}}$ gives repulsion between neighbouring doublons and holes. In \eqr{Eqn:UndrivenEffectiveHam} these terms have an identical coupling given by super-exchange interaction $J_0 = 4t^2/U$.

The isotropy between $H_{\text{hop}}$ and $H_{\text{int}}$ makes the ground state of $H_{\text{eff}}$ with $N$ doublons a so-called $\eta$-pair state $\ket{\eta_N} \sim (\eta^+)^N\vac$, where $\eta^+ \sim \sum_i(-1)^i d^\dagger_i$ creates a doublon at $q=\pm \pi/a$ momentum. This is a direct manifestation in $H_{\text{eff}}$ of the celebrated $\eta$-SU(2) symmetry of the Hubbard model, $\comm{H_{\text{hub}}}{\eta^+} = U{\eta^+}$, first introduced by Yang~\cite{Yang1989}. Consequently, $\ket{\eta_N}$ is an {\em exact} eigenstate of $H_{\text{hub}}$ with an energy $E = UN$ for any $U$. Furthermore, since
    \begin{align} 
        \bra{\eta_N}d^\dagger_i d_j\ket{\eta_N} = \frac{N(L-N)}{L(L-1)}e^{\mi \pi(i-j)},
    \end{align}
the $\eta$-pair state displays staggered off-diagonal long-range order consistent with $N$ doublons Bose condensing at the zone edge $q_{\rm c} = \pm\pi/a$.  Various proposals for generating in the Hubbard model an $\eta$-condensate, or states with $\eta$-like correlations, have been put forward. These include adiabatic switching of an optical lattice confinement and superlattice potentials \cite{Kantian2010}, flipping of the band structure induced by driving the attractive Hubbard model \cite{Kitamura2016}, and as an eigenstate of a ``dark" Hamiltonian created from a Hubbard model with spin dephasing \cite{Berislav2019}. 
 
\subsection{Generalizing to driven system} \label{sec:effective_driven_model}
For $\tau \geq 0$ the driven Hubbard model in \eqr{EQN:Driven_Hubbard_Model} is time-periodic $H_\text{full}(\tau) = H_\text{full}(\tau + T)$. Analogous to Bloch's theorem for discrete spatial translational symmetry, discrete
time translational symmetry constrains the solutions of the time-dependent Schr{\"o}dinger equation via Floquet's theorem \cite{Eckardt2015}. The key result is that the time-evolution operator $\mathcal{U}(\tau_2,\tau_1)$ of the driven system between two times $\tau_1$ and $\tau_2$ can be decomposed into a product of three unitaries
    \begin{align} 
        \mathcal{U}(\tau_2,\tau_2) = e^{-\mi K(\tau_2)} e^{-\mi(\tau_2 - \tau_1)H_F}e^{\mi K(\tau_1)}.
    \end{align}
Here $H_F$ is the time-independent Floquet effective Hamiltonian which generates continuous evolution between $\tau_1$ and $\tau_2$. This is sandwiched between two unitaries generated by the hermitian kick operator $K(\tau)$, with parametric dependence on the application time $\tau$ such that $K(\tau) = K(\tau + T)$. The kicks describe micromotion within each drive period. We will discuss its impact on the observables in \secr{sec:cold_atom_setup}. In particular we find that the initial kick has no effect on the density or momentum distributions, meaning the effective Hamiltonian $H_F$ is sufficient to fully characterise the dynamics at times $\tau$ that are integer multiples of $T$, namely the \emph{stroboscopic} time evolution of the driven system.

In general the computation of $H_F$ is a highly non-trivial problem. As detailed in Appendix \ref{sec:app_floquet}, we use an approach based on diagonalizing the so-called Floquet quasi-energy operator $F$ in an extended Floquet-Hilbert space after transforming to the frame rotating with the driving via the unitary $R(\tau) = \exp[-\mi \int d \tau' H_{\text{drv}}(\tau')]$. Specifically, for the driven Hubbard model, $F$ is composed of diagonal blocks that are replicas of the Hubbard Hamiltonian $H_\text{hub}$, each shifted in energy by $\Omega$ relative to its neighbouring block, and with $H_\text{kin}$ renormalized by the drive by $\mathcal{J}_0(\nu)$, where $\nu = A/\Omega$ and $\mathcal{J}_m(\nu)$ is the $m$th Bessel function of the first kind. The off-diagonal couplings between blocks $m$ and $m'$ is given by $H_\text{kin}$ renormalized by $\mathcal{J}_{m'-m}(\nu)$, corresponding to fermion hopping accompanied by an exchange of $m'- m$ quanta with the drive. 

\begin{figure}[h]
\centering
\includegraphics{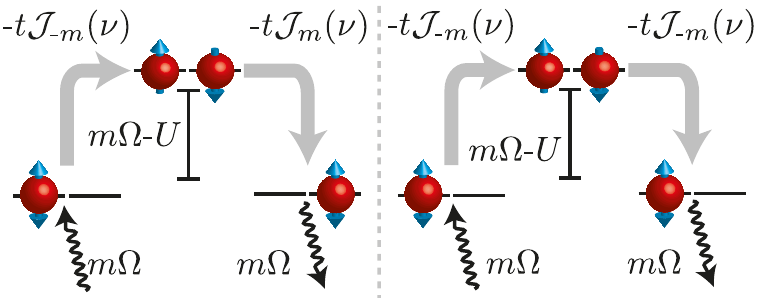}
\caption{A depiction of the second order processes contributing to (a) $J(\nu,\Omega)$ in \eqr{Eqn:DrivenSuperExchange} and (b) $\Delta(\nu,\Omega)$ in \eqr{Eqn:DrivenAnisotropy}. \label{FIG:Floquet_Process}}
\end{figure}

The diagonalization of $F$ is accomplished approximately by again employing standard degenerate perturbation theory in which we isolate simultaneously the band of highest-energy eigenstates of $H_\text{hub}$ within one block and account for corrections due to virtual transitions to eigenstates both in this block and all others. To second order in $t/U$ we obtain a driving dependent effective Hamiltonian  
\begin{align}\label{Eqn:DrivenEffectiveHam}
	\begin{split}
	\tilde{H}_{\text{eff}}(\nu,\Omega) = \mathbbm{P}\left[\frac{J(\nu,\Omega)}{2}\Big(H_{\text{hop}} + \Delta\left( \nu,\Omega \right)H_{\text{rep}}\Big)\right] \mathbbm{P},
	\end{split}
\end{align}
where the modified super-exchange and anisotropy are
\begin{align}\label{Eqn:DrivenSuperExchange}
	J(\nu,\Omega) = J_0 \sum_{m = - \infty}^{\infty} (-1)^m \frac{\mathcal{J}_m(\nu)^2}{1+m\Omega /U },\\ 
	\Delta(\nu,\Omega) = \frac{J_0}{J(\nu,\Omega)} \sum_{m = - \infty}^{\infty} \frac{\mathcal{J}_m(\nu)^2}{1+m\Omega /U}, \label{Eqn:DrivenAnisotropy}
\end{align}
obtained by summing over all blocks $m$.  

\begin{figure}[t]
\centering
\includegraphics{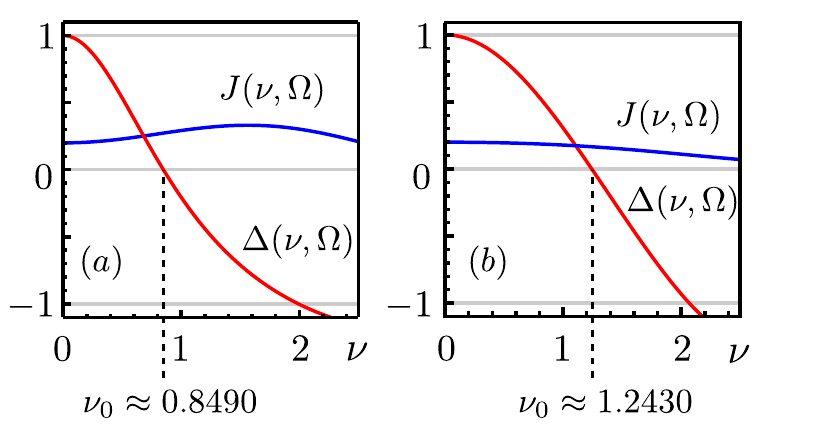}
\caption{Parameters $J$ and $\Delta$ of the effective spin model as a function of dimensionless driving strength $\nu = A/\Omega$, $U = 20t$, driving frequency is $\Omega = 24t$ in (a) and $\Omega = 30t$ in (b). The marked points $v_0$ indicate the dimensionless driving strength at which the model is rendered non-interacting.} 
\label{FIG:Effective_J_Delta}
\end{figure}

 The $(-1)^m$ factor in $J(\nu,\Omega)$ arises because the hopping of a doublon consists of two single particle hops in the same direction, as in \fir{FIG:Floquet_Process}(a). If the first hop is Fourier shifted by $m\Omega$ with an amplitude $\mathcal{J}_m(\nu)$ then the second hop returning to the same initial band would be a Fourier shift of $-m\Omega$ with an amplitude $\mathcal{J}_{-m}(\nu) = (-1)^m \mathcal{J}_{m}(\nu)$. In contrast the doublon-hole repulsion strength arises from the hopping of a single fermion forwards and back, with the same amplitude $\mathcal{J}_m(\nu)$, as in \fir{FIG:Floquet_Process}(b). Consequently, the effective model has a driving induced anisotropy that breaks the $\eta$-SU(2) symmetry of the undriven model \cite{Kitamura2016,Nocera2017}. The reason for this is that the effective model describes charge degrees of freedom and the driving couples directly to charge. Had we considered instead the more conventional lowest-lying eigenstates of the half-filled Hubbard model, where the effective model describes spin degrees of freedom, no such anisotropy would be induced since the driving does not break the spin SU(2) symmetry. 

In the undriven limit we have $\lim_{\nu \rightarrow 0}~J(\nu,\Omega) = J_0$ and $\lim_{\nu \rightarrow 0}~\Delta(\nu,\Omega) = 1$, recovering \eqr{Eqn:UndrivenEffectiveHam}. For any finite $\nu$ in the high-frequency driving limit, $\Omega \gg U \gg t$, only the $m=0$ contribution to \eqr{Eqn:DrivenSuperExchange} and \eqr{Eqn:DrivenAnisotropy} survives giving $\lim_{\Omega/U \rightarrow \infty}~J(\nu,\Omega) = J_0 \mathcal{J}_0(\nu)^2$ and $\lim_{\Omega/U \rightarrow \infty}~\Delta(\nu,\Omega) = 1$, so isotropy is preserved with a renormalized super-exchange. Thus $\eta$-SU(2) symmetry breaking arises when $\Omega/U$ is finite. Here we will study driving frequencies $\Omega \gtrsim U$, while avoiding direct resonant couplings between Hubbard excitations to first order in $t$ \cite{Bukov2016a}.

In \fir{FIG:Effective_J_Delta} we plot $J(\nu,\Omega)$ and $\Delta(\nu,\Omega)$ for $U=20t$ for two moderate driving frequencies. In both cases the anisotropy $\Delta(\nu,\Omega)$ is suppressed for moderate non-zero amplitudes $\nu$. The doublon-hole repulsion can even be completely removed at $\nu_0$, as highlighted in \fir{FIG:Effective_J_Delta}, where the doublons then behave as non-interacting hard-core bosons. For both $\Omega$'s stronger driving can induce doublon-hole attraction $\Delta(\nu,\Omega)<0$.

\subsection{Doublon domain melting}
\label{sec:doublon_melting}

\begin{figure}[t]
	\centering
	\includegraphics{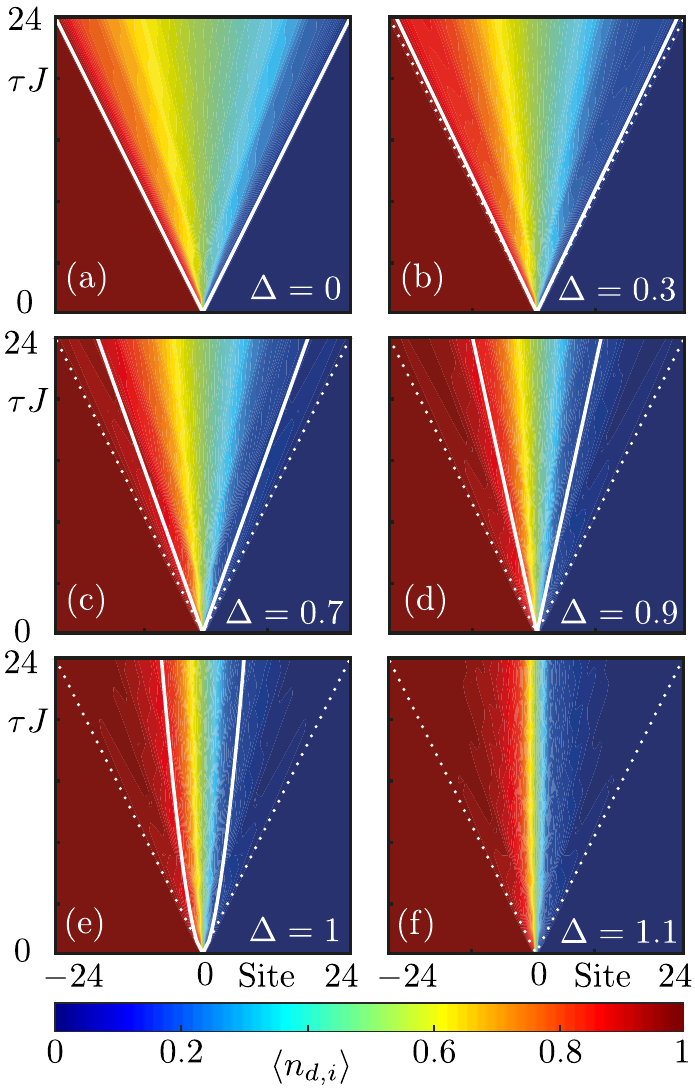} 
	\caption{Time evolution of the site dependent magnetisation as the domain wall melts for various interaction strengths. The outer dashed lines represent the maximal velocity $J$. The inner solid lines in (b),(c) and (d) indicate the slower melting of the domain with a velocity $J\sqrt{1-\Delta^2}<J$. In (e), the inner curve represents the super-diffusive nature of the isotropic point $\Delta = 1$, with the domain wall spreading as a power law in time with an exponent $\approx 3/5$. Finally, in (f) for $\Delta>1$ after some initial transients the melting remains localized and has neither ballistic nor super-diffusive behaviour. These calculations used a lattice size $L=60$, with $N = 25$ particles, time step $J\delta\tau = 0.01$ and an MPS bond dimension $\chi = 1200$.}
	\label{FIG:Spin_Melting_Panel}
\end{figure}

\begin{figure}[t]
	\centering
	\includegraphics{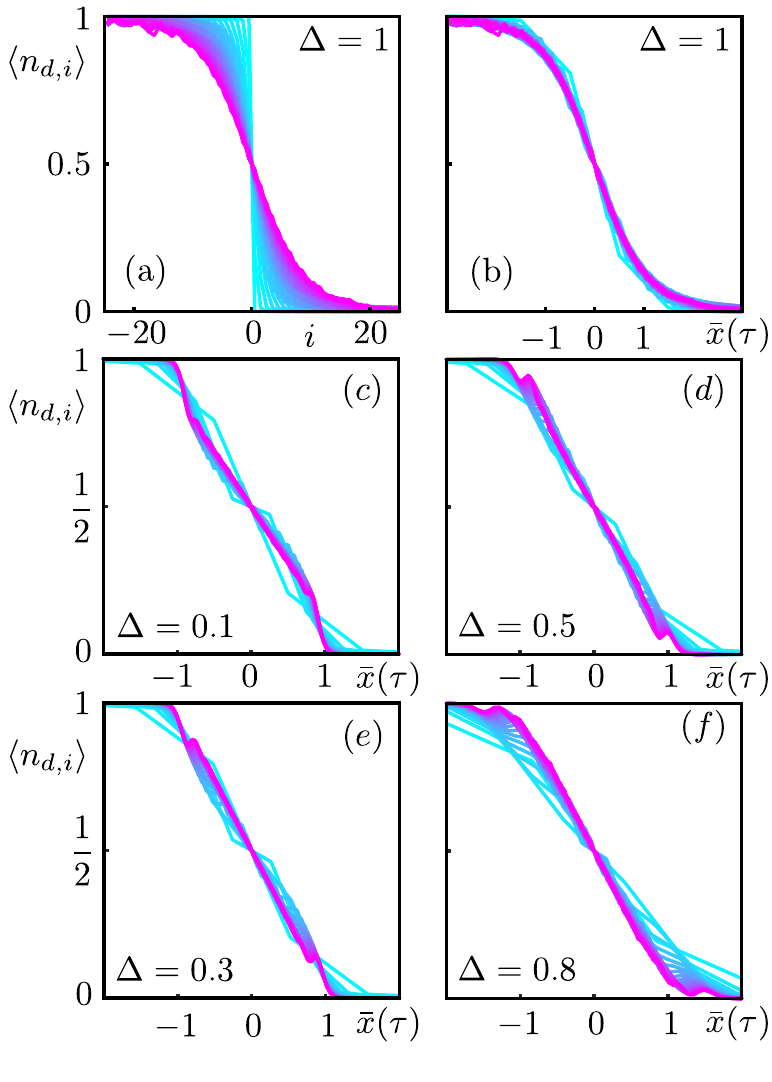} 
    \caption{(a) For a system of size $L=60$ with $N=25$ particles $\av{d^\dagger_i d_i}$ is shown for times $J\tau = 0,2,..24$ around the domain wall boundary. (b) The same plot of density $\av{d^\dagger_i d_i}$ but with the site coordinate $i$ rescaled as $\bar{x}(\tau)= a(i-N)/ \sqrt{\tau}$ with the collapse indicating a universal form is approached. (c)-(f) Show the collapse of melt profiles for $\Delta < 1$ using a rescaling $\bar{x}(\tau) = a(i-N)/v(\Delta) \tau$. Same numerical details as \fir{FIG:Spin_Melting_Panel}}
	\label{FIG:Rescaled_Melt_Front}
\end{figure}

We have established that in the regime $U \gg t$ and $\Omega \gtrsim U$ the stroboscopic dynamics of the periodically driven Hubbard model introduced in \secr{sec:model} reduces to an effective model. Specifically, the driven dynamics is approximated as a sudden quench at $\tau \geq 0$ of interacting hard-core bosons governed by the effective model in \eqr{Eqn:DrivenEffectiveHam} specified by $J(\nu,\Omega)$ and $\Delta(\nu,\Omega)$ with an initial state $\ket{11 \dots 1100 \dots 00}$. Analysing this effective model brings several benefits. First, it is computationally much simpler than the full Hubbard model, allowing much larger system sizes to be accessed numerically. Second, it is isomorphic to magnetic domain-wall melting in the XXZ spin model, so insight can be gleaned from the extensive studies on this spin model \cite{Sabetta2013,Collura2018,Misguich2017} .

For $0 \leq \Delta \leq 1$ the quench dynamics of the initially sharp domain generically displays an expansion of bosons outwards from the boundary of the domain, and correspondingly holes inwards into the domain. For a selection of $\Delta$'s, \fir{FIG:Spin_Melting_Panel} shows the evolution of the doublon density profile $\av{n_{d,i}(\tau)}$ up to a time $\tau = 24/J$. Centred on the domain boundary we see that a melt lobe forms with a size that grows linearly in time for $0\leq \Delta <1$, grows sub-linearly for $\Delta = 1$ and stops growing after finite time for $\Delta>1$. For $0 \leq \Delta \leq 1$ the melt profile displays a universal form. Formally this is revealed by rescaling the site coordinates in some appropriate way $\bar{x}(\tau)$ such that the melt profile at all times collapses onto itself. To illustrate this \fir{FIG:Rescaled_Melt_Front}(a) shows the melt profile for each site $i$ for times between $\tau = 5/J$ and $\tau = 24/J$ with $\Delta = 1$. Owing to the anomalous super-diffusive transport properties at the isotropic point a power-law rescaling $\bar{x}(\tau) = a(i - N)/ (J \tau)^p$ with $p = 3/5$ \cite{Misguich2017} is found to induce the profile collapse, as shown in \fir{FIG:Rescaled_Melt_Front}(b). For all other interaction strengths $0 \leq \Delta < 1$, a linear rescaling $\bar{x}(\tau) = a(i-N)/v(\Delta) \tau$ with a $\Delta$ dependent speed $v(\Delta) = J\sqrt{1-\Delta^2}$ establishes universality, as depicted in \fir{FIG:Rescaled_Melt_Front}(c)-(f). For $\Delta = 0$ the density profile in the universal region is known \cite{Antal1999} to be $\av{n_{d,i}(\tau)} = \arccos(\bar{x}(\tau))/\pi$.

\begin{figure}[t]
	\centering
	\includegraphics[scale=0.9]{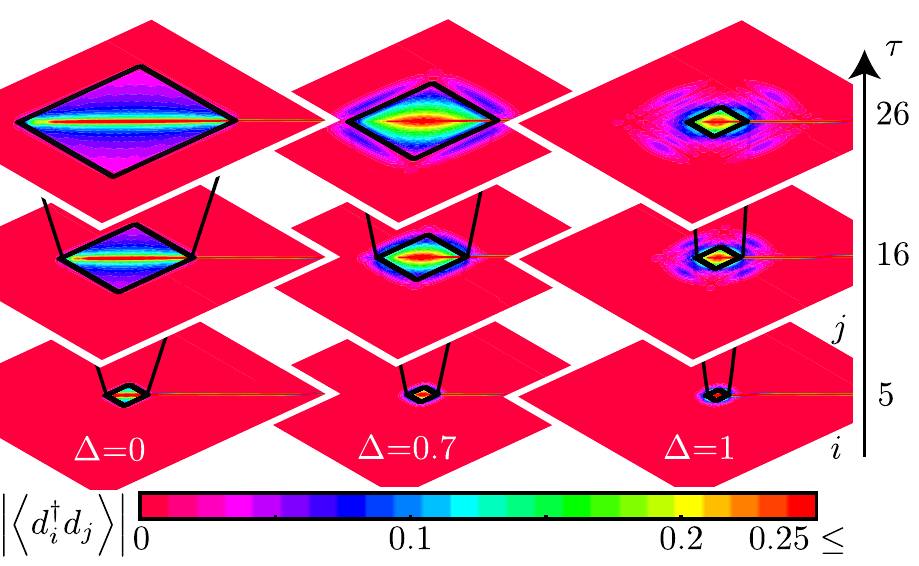}
    \caption{(a) For a system of size $L=100$ with $N=50$ particles a colormap around the domain wall boundary of the OPDM $\av{d^\dagger_i d_j}$ is shown for three time slices $J\tau = 5,16,26$. The expanding black square delineates the universal region. The same MPS bond dimension $\chi = 1200$ and time step $J\delta\tau = 0.01$ was used.  }
	\label{FIG:OPDM_3D}
\end{figure}

Here our central focus is on the properties of the system, beyond the density profile, within the melt lobe universality region $-1 \leq \bar{x}(\tau) \leq 1$ as a function of the interaction $0 \leq \Delta \leq 1$. In particular, as first discovered by Rigol and Muramatsu \cite{Rigol2004}, for quenches in the non-interacting $\Delta = 0$ limit dynamical quasicondensation occurs. This novel effect is only revealed by examining the full doublon one-particle density matrix (OPDM)
\begin{align}
  \rho_{ij}(\tau) =  \bra{\phi(\tau)} d^\dagger_i d_j \ket{\phi(\tau)},
 \end{align}
computed from the time-evolving state $\ket{\phi(\tau)}$ of the effective model. In \fir{FIG:OPDM_3D} a colormap of the OPDM for some representative interaction strengths are shown for three time slices. The expanding black square signifies the universal region. For $\Delta = 0$ the universal region expands at the maximum speed $J$ of the single-doublon band and so spans the light-cone of the system dynamics. Even with interactions there remains small contributions to the OPDM throughout the light-cone due a tiny fraction of doublons escaping the domain at a speed $J$ from short-time perturbative adjustments to the sudden quench. However, for $\Delta >0$ the expansion speed $v(\Delta)$ defining the melt lobe becomes increasingly slow in comparison. As we shall show this separation of speeds makes the analysis of the interacting system numerically challenging because the full light-cone must still be captured such that $L \sim N \sim J\tau$ to avoid spurious boundary effects.

By exploiting the equivalence of the $\Delta = 0$ limit to non-interacting spinless fermions Rigol and Muramatsu \cite{Rigol2004} solved numerically exactly the dynamics of large systems over long times. Surprisingly, they found that the evolution of the momentum distribution of the hard-core bosons
\begin{align}
n_q(\tau) = \frac{1}{L}\sum_{ij} e^{-\mi q \left( i-j  \right) a} \rho_{ij}(\tau),
\label{EQN:Momentum_Expectation}
\end{align}
quickly changes from a featureless constant, characterising the localized state at $\tau = 0$, to a distribution with a pronounced peak at finite-momentum $q_{\rm c}a = \pi/2$, indicative of a condensate forming dynamically. This result is reproduced here in \fir{FIG:Momentum_Distributions}(a) using td-DMRG. 

Our analysis of the interacting system similarly begins by examining the momentum distribution. A simple energetic argument anticipates that a peak in the momentum will shift towards $qa = \pi$ as $\Delta$ approaches the isotropic point, and that no peak will develop for $\Delta >1$. Initially, the sharp domain wall has an average energy $\av{H_\text{eff}} = J\Delta/2$ coming exclusively from the doublon-hole repulsion at the interface $...1 \ 1 \ 1 \ 1 \ 0 \ 0 \ 0 \ 0 \ ...$. If we now consider a doublon at the interface unbinding and propagating into the empty region as $...1 \ 1 \ 1 \ 0 \ 0 \ 1 \ 0 \ 0 \ ...$ then the interfacial interaction energy remains unchanged, but the isolated doublon now contributes an additional interaction energy of $J\Delta$. If this melting is to occur conservation of energy demands that the increased interaction is compensated by a reduction in the isolated doublon's kinetic energy. Given that the single-doublon dispersion is $\epsilon_{\text{hop}}(k) = J\cos(qa)$ this implies that the propagating doublon will possess a momentum 
\begin{align}
 q_{\rm c}a \sim \pm \arccos(-\Delta) =  \pm \arccos(\Delta) + \pi. \label{EQN:mom_shift}
\end{align}
Assuming the melt lobe is dilute in the long-time limit we thus anticipate the accumulation of melting doublons into this momentum state. The relation \eqr{EQN:mom_shift} is in agreement with $q_{\rm c}a = \pi/2$ for $\Delta =0$, and also reveals that for $\Delta > 1$ the finite bandwidth is insufficient to compensate the interaction precluding condensation at any $q_{\rm c}$.

To confirm these expectations we have computed the momentum distribution from the interacting OPDM. Crucially, to isolate the contributions from the melting, \eqr{EQN:Momentum_Expectation} was not applied directly. Instead, we first restricted the Fourier transform of the OPDM to the universal melt lobe region defining a subsystem centred on the domain boundary with a time-dependent size $\ell(\tau) = 2v(\Delta)\tau/a$ for $0 \leq \Delta < 1$, or $\ell(\tau) = 2(J\tau)^p/a$ for $\Delta =1$ \footnote{To smooth the integer jumps in the melt lobe size we performed weighted averages between adjacent sizes.}. Next, to allow a clean comparison to \eqr{EQN:mom_shift} given the limited resolution in momentum space available for small systems, we applied a phase-shift to the OPDM as
\begin{align}
  \rho_{ij}(\tau) \rightarrow \rho_{ij}(\tau)e^{\mi (i-j)\arccos(\Delta)},
\end{align}
before taking the Fourier transform. The purpose of this was to shift any contribution at $q_{\rm c}$ to precisely $q = \pi/a$, which is guaranteed to coincide with a discrete momenta available since $\ell(\tau)$ is always even. After taking the Fourier transform the momentum grid was then shifted back. The resulting $n_q$ are shown in \fir{FIG:Momentum_Distributions} for several interaction strengths and times. The number of doublons in the melt lobe is $N(\tau) = \ell(\tau)/2$.

 \begin{figure}[t]
 	\centering
 	\includegraphics{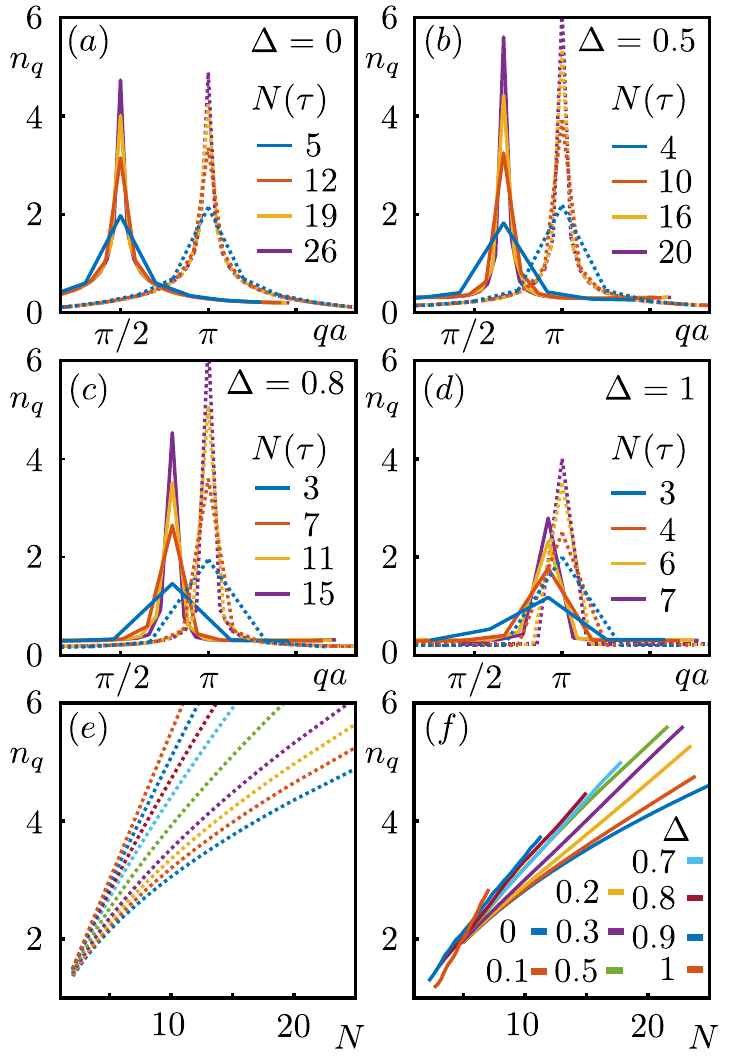}
     \caption{ The momentum distributions within the universal region of the melt state plotted for a sequence of increasing occupations (or times) $N(\tau) = \ell(\tau)/2$. Also plotted (dotted lines) is the $qa = \pi$ momentum peak of the ground state of the same Hamiltonian for the same sequence of doublon numbers at half filling. Calculation parameters are identical to those of \fir{FIG:OPDM_3D}.}
 	\label{FIG:Momentum_Distributions}
 \end{figure}

The evolution of the momentum distributions in \fir{FIG:Momentum_Distributions}(b)-(d) reveal an intriguing parallel to the non-interacting behaviour in \fir{FIG:Momentum_Distributions}(a) -- as time progresses we see the emergence of an increasingly sharp peak at finite momentum close to $q_{\rm c}$ predicted by \eqr{EQN:mom_shift}. These peaks are a smoking gun of dynamical quasi-condensation, and indicate that this effect is not simply an anomalous feature of the non-interacting system. Also plotted are the momentum distributions of the half-filled ground state calculated with DMRG using the same interaction strength $\Delta$ and system size $\ell(\tau)$ as the corresponding melt state. All ground states display a peak at $qa = \pi$ that becomes sharper with increasing $\Delta$ and whose magnitude grows with $N$ sub-linearly for $0 \leq \Delta < 1,$ consistent with quasi-condensation, and linearly for $\Delta =1$ consistent with $\eta$-condensation, as shown in \fir{FIG:Momentum_Distributions}(e). 

The behaviour of the melt state peak at $q_{\rm c}$ in \fir{FIG:Momentum_Distributions}(f) displays the same trends. While the ground state peak growth with $N$ settles quickly into it a universal asymptotic form the melt dynamics with interactions lags behind and the form of its growth form has yet to fully stabilise. Nonetheless for $\Delta < 0.5$ there is good correspondence between the melt state and ground state peak growth. For $\Delta > 0.5$, even though there are severe melt lobe size (or melt time) limitations, the growth appears suppressed. For $\Delta = 1$ the peak in the melt state distribution shown in \fir{FIG:Momentum_Distributions}(d) does not occur at exactly at $qa = \pi$, but is instead located at the discrete momentum state directly adjacent to it, suggesting it may approach $qa = \pi$ only in the thermodynamic limit. For $\Delta >1$ (not shown) a small peak emerges, caused by the sharp initial state transients, whose magnitude rapidly saturates with $N(\tau)$.      

For $\Delta = 0$ it is known that the melt state asymptotically converges to a boosted ground state~\cite{Vidmar2017}, apparent here already for relatively short times with the similarity of the peaks in \fir{FIG:Momentum_Distributions}(a). A natural question is whether melting gives a boosted ground state more generally in the interacting case. An affirmative answer was obtained in previous work which focused on a initial ``soft'' domain wall \cite{Sabetta2013,Lancaster2010}. In this case the initial state was the ground state of a weak confinement potential $ v_i \propto \tanh(\beta i)$, where the constant $\beta$ controls the wall width. As a result the initial state has a small difference in the density $\delta \ll 0.5$ far to the left $\av{n_{d,i\ll0}} \sim 0.5 + \delta$ and far to the right $\av{n_{d,i\gg0}} \sim 0.5 - \delta$ of the interface. Using both numerical~\cite{Sabetta2013} and analytic hydrodynamic arguments~\cite{Lancaster2010} it was shown that the correlations of the melt state in the long-time limit generated around the domain boundary (far from the edges of the chain) have a simple relation to the corresponding half-filled ground state as
\begin{align}
  \av{d^\dagger_i d_j}_\text{melt}  =  \av{d^\dagger_i d_j}_{\text{gs}}~e^{-\mi \theta (\di - j)}, \label{EQN:gs_corr}
\end{align}
where $\theta = 2\delta \arccos(\Delta)$. For the setup we consider here, where the ground states displays a peak at $qa=\pi$ and a ``hard'' domain wall is used so $\delta = 0.5$, we see that this phase relation agrees with \eqr{EQN:mom_shift} used above. Yet there are some differences in the behaviour of the melt states and ground states shown in \fir{FIG:Momentum_Distributions}. To better assess these difference we now move on to compute other properties of the melt state expected of a quasicondensate. 

\subsection{Further signatures of quasicondensation} \label{sec:quasicondense_signatures}
While the emergence of a sharp peak in the momentum distribution is suggestive of a condensation phenomenon a more careful examination of the scaling properties of the OPDM is needed to fully identify the nature of the melt state. Following the analysis for $\Delta = 0$ performed by Rigol and Muramatsu~\cite{Rigol2004} we diagonalized the OPDM to find the natural orbitals and their occupations as a function of time after the quench. The emergence of quasicondensation is revealed by the behaviour of the so-called lowest natural orbital (LNO) $\phi_0$ with the largest occupation $\lambda_0$. For simple unfragmented dynamical quasicondensation the LNO should display, after transients have subsided, a time-invariant universal form with the rescaled of coordinate $\bar{x}(\tau)$. In \fir{FIG:LNO_Shape} the rescaled LNO's for a variety of $\Delta$'s are shown for times $15 \leq J\tau \leq 25$. Identically to the density profile we find that within the melt lobe $|\bar{x}(\tau)| \leq 1$ the form of LNO for all $\Delta$'s is effectively constant in time confirming its universality. Decaying contributions outside the melt lobe are seen in all cases, but are especially prominent for larger interactions due to the slower expansion speed.

\begin{figure}[t]
	\centering
	\includegraphics{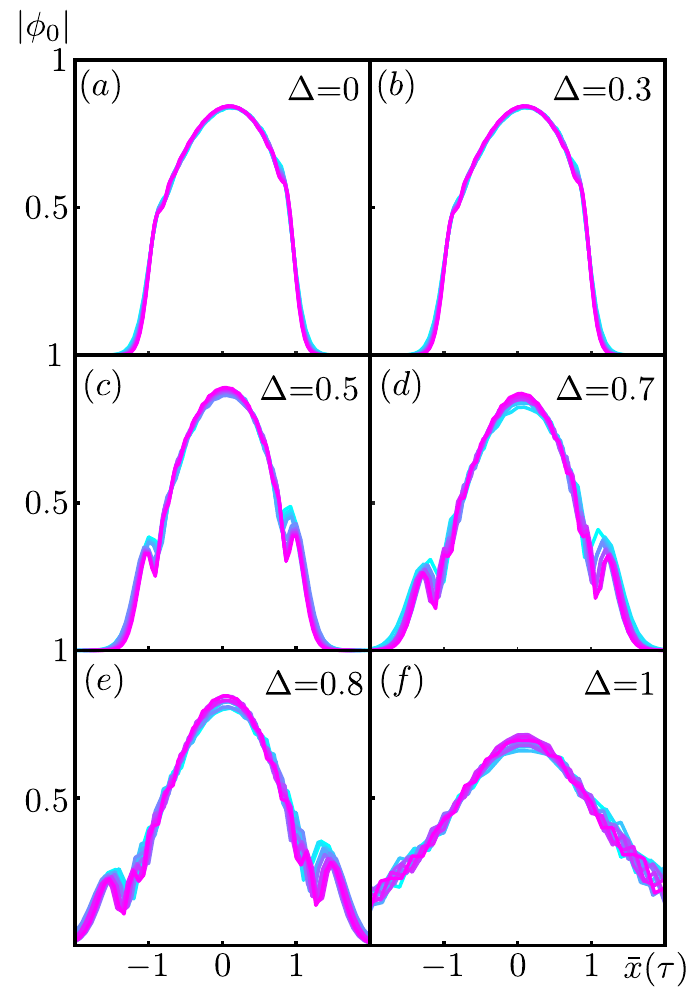}
	\caption{ The form of the absolute value of the natural orbitals as a function of $\bar{x}(\tau)$ plotted for times $\tau = 15,16, \dots ,25$. Within $-1 < \bar{x}(\tau) <1$, $|\phi_{0}|$  largely remains constant. The most significant change with $\Delta$ is the enlarged tails caused by single doublons moving off from the more slowly expanding melt region. Calculation details identical to \fir{FIG:OPDM_3D}} 
	\label{FIG:LNO_Shape}
\end{figure}

The next crucial property for quasicondensation is the scaling of the LNO occupation $\lambda_0$ with the number of doublons $N(\tau)$ in the melt lobe. True condensation occurs if $\lambda_0(N) = \alpha N$, implying a finite O(1) fraction of particles $\sigma(N) = \lambda_0(N)/N$ occupy the LNO in the thermodynamic limit. Quasicondensation is instead a sub-linear power law growth of $\lambda_0(N) \propto N^p$ with $0<p<1$ so the occupancy fraction vanishes as $\sigma(N) \sim 1/N^{1-p}$. Analogous to our analysis of the momentum distribution we determined LNO's by diagonalizing the OPDM only inside the melt lobe region $-1<\bar{x}(\tau) < 1$ for each $\Delta$. The behaviour of $\sigma(N)$ for the corresponding half-filled ground state for the same $\Delta$ and particle number $N(\tau)$ is shown in \fir{FIG:LNO_Growth}(a). For $0 \leq \Delta < 1$ this shows a power-law decrease in $\sigma(N)$ with $N$, but the decay slowing down with increasing $\Delta$. At the point $\Delta = 1$ the LNO density saturates to a constant $\sigma(N) \rightarrow 0.5$ signalling $\eta$-condensation. 

In \fir{FIG:LNO_Growth}(b) we show $\sigma(N)$ for the melt state. For $\Delta = 0$ a power-law decay identical to that of the ground state is seen. For interacting systems $\Delta > 0$ the basic trend is similar to the ground state with the decay in $\sigma(N)$ slower than $\Delta = 0$ and softening with increasing $\Delta$. This is indicative of stronger dynamical quasicondensation with increasing interactions, consistent with the sharpening momentum distributions with $\Delta$ observed already in \fir{FIG:Momentum_Distributions}(a)-(d). However, for all interacting cases the behaviour of $\sigma(N)$ displays more discernible differences to the ground state than was apparent in the momentum distribution alone. None of the decay curves have reliably converged to a power-law form, even for the weakest interactions, and there are even signs of saturation. However, due to the limitation in reachable $N(\tau)$'s, which is most severe as $\Delta \rightarrow 1$, it is currently inconclusive whether genuine dynamical {\em condensation} is occurring.

\begin{figure}[t]
	\centering
	\includegraphics{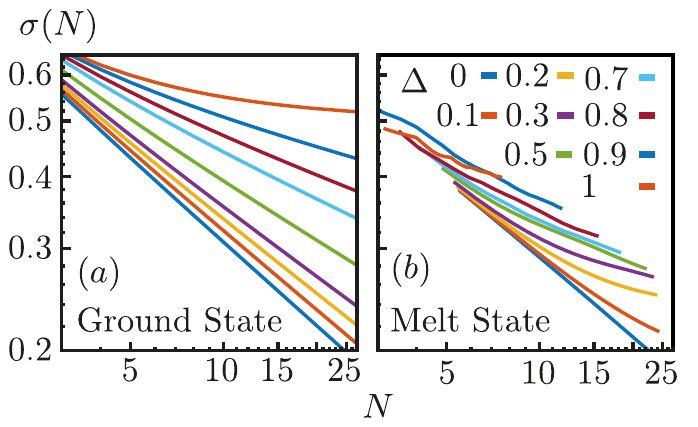}
	\caption{ The density of the LNO occupation as a function of particle number $\sigma(N) = \lambda_0(N)/N$ for (a) the ground state (b) the state during the melting. A monotonic increase in $\sigma(N)$ is observed for increased $\Delta$, but also a shrinking range of $N(\tau)$ is accessible numerically. Calculation details identical to \fir{FIG:OPDM_3D}.}
	\label{FIG:LNO_Growth}
\end{figure}

Another important property of quasicondensation is the decay of long-ranged correlations off-diagonal correlations in the OPDM within the spatial support of the LNO. For the non-interacting case a distinctive power-law decay is found
\begin{align}
 \rho_{ij} \propto \frac{e^{\mi q_{\rm c} (i-j)a}}{\sqrt{|i-j|}},
\end{align}
with a phase difference of $q_{\rm c}a = \pi/2$ between neighbouring sites signalling quasicondensation at finite momentum \cite{Rigol2004b}. The observed $1/2$ exponent for the algebraic decay of correlations is identical to the ground state of the same non-interacting system, as expected from \eqr{EQN:gs_corr}. In \fir{FIG:Off_Diagonals} the absolute value of doublon correlations $|\rho_{N,j}|$ from the domain wall boundary at a time $J\tau = 30$ are shown for various $\Delta$'s. Here for all non-zero interactions we see even more visible differences from the power-law decay of the corresponding ground state that are also shown. For $\Delta \leq 0.5$ in \fir{FIG:Off_Diagonals}(a)-(c) the melt state displays stronger correlations over the melt lobe than the ground state, while in contrast for $\Delta > 0.5$ in \fir{FIG:Off_Diagonals}(d)-(f) they are increasingly weaker. None of the interacting cases considered over the times accessible in our calculations agree with the soft wall hydrodynamic prediction \footnote{We cannot preclude the possibility that calculations on much longer timescales, not currently feasible with td-DMRG for this hard domain wall setup, might recover the behaviour given in \eqr{EQN:gs_corr} as the interface softens.} in \eqr{EQN:gs_corr}.

Overall, we have found distinctive signatures of dynamical quasicondensation within the effective model of doublons arsing from the driven Hubbard model. However, due to the separation of expansion speeds the system sizes reached in our calculations are insufficient to fully quantitatively diagnose the nature of this quasicondensation.     

\begin{figure}[t]
	\includegraphics[scale=0.9]{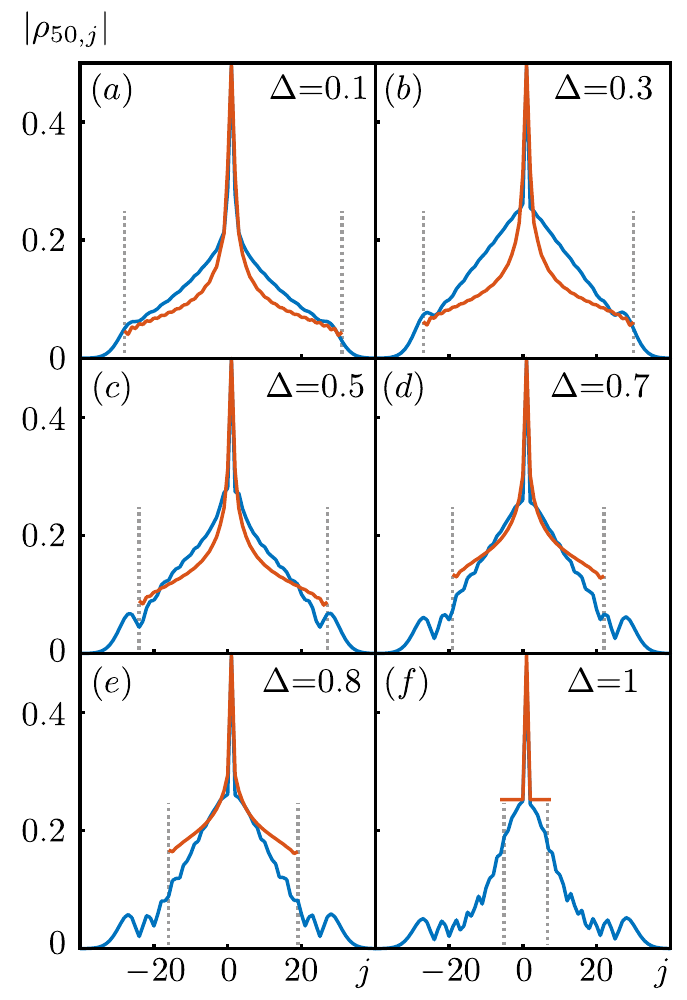}
	\caption{ The absolute value of the off-diagonal correlations of the OPDM $\rho_{ij}$ for fixed $i=N=50$ at the domain wall and varying $j$. Solid lines is the off diagonals for the melted domain wall plotted at a time $\tau J = 30$ for a selection of $\Delta$'s. Dotted lines (black) are the correlations of the ground state of a system with the same Hamiltonian and system size at half filling. The dotted vertical lines delineate the melt lobe region which shrink rapidly as $\Delta \rightarrow 1$. Calculation details idential to \fir{FIG:OPDM_3D}.}
	\label{FIG:Off_Diagonals}
\end{figure}

\section{Driven Hubbard Model} \label{sec:driven_full_system}
Having observed that a form of quasicondensation emerges within the effective model we now return to the full driven Hubbard model in \eqr{EQN:Driven_Hubbard_Model}. In particular we will now demonstrate that this novel effect is robust beyond the strongly interacting and high-frequency approximations underlying the validity of the effective model. Moreover, we will show that it continues to occur for large but finite interactions $U > t$ and for finite driving frequencies $\Omega > U$ not too close to resonance.  

\subsection{Zero driving case} \label{sec:undriven_hubbard}
As a baseline we consider first the undriven Hubbard model, corresponding to $\Delta =1$ in the effective model, and examine agreement as $U/t$ is decreased. A readily accessible indicator within td-DMRG of the increased complexity of the full Hubbard model is the entanglement entropy of the time-evolved state $\ket{\psi(\tau)}$
\begin{align}
 S_E(i) =  -\sum_\alpha \Lambda^{[i]}_\alpha \log\left(\Lambda^{[i]}_\alpha\right),
\end{align}
where $\Lambda^{[i]}_\alpha$ are the squared Schmidt coefficients of $\ket{\psi(\tau)}$ for a bipartition of the system between sites $i$ and $i+1$. In \fir{FIG:Undriven_Doublon_Num_and_LNO}(a) $S_E(i)$ is plotted for various $U/t$'s for sites $i$ close to the domain wall after a time $J\tau = 10$ with $J = 2t^2/U$. For this short time the slow expansion for $\Delta = 1$ gives a melt lobe extending around $ \sqrt{10}  \sim 3$ sites either side of the wall interface. As expected the strongest interaction $U/t = 20$ closely follows the $S_E(i)$ of the effective model up to the speed $J$ light-cone of 10 sites, and so captures the small contribution of the ballistically escaping doublons. However, the full Hubbard model solution also has a non-negligible entropy beyond the effective model's light-cone arising from the partial disassociation of doublons into fast-moving fermions with a speed $2t>J$. The full Hubbard model thus has yet another speed associated to the quench dynamics. As expected this fermion contribution becomes more pronounced with decreasing $U/t$ as doublons become less stable. 

\begin{figure}
	\includegraphics[scale=0.9]{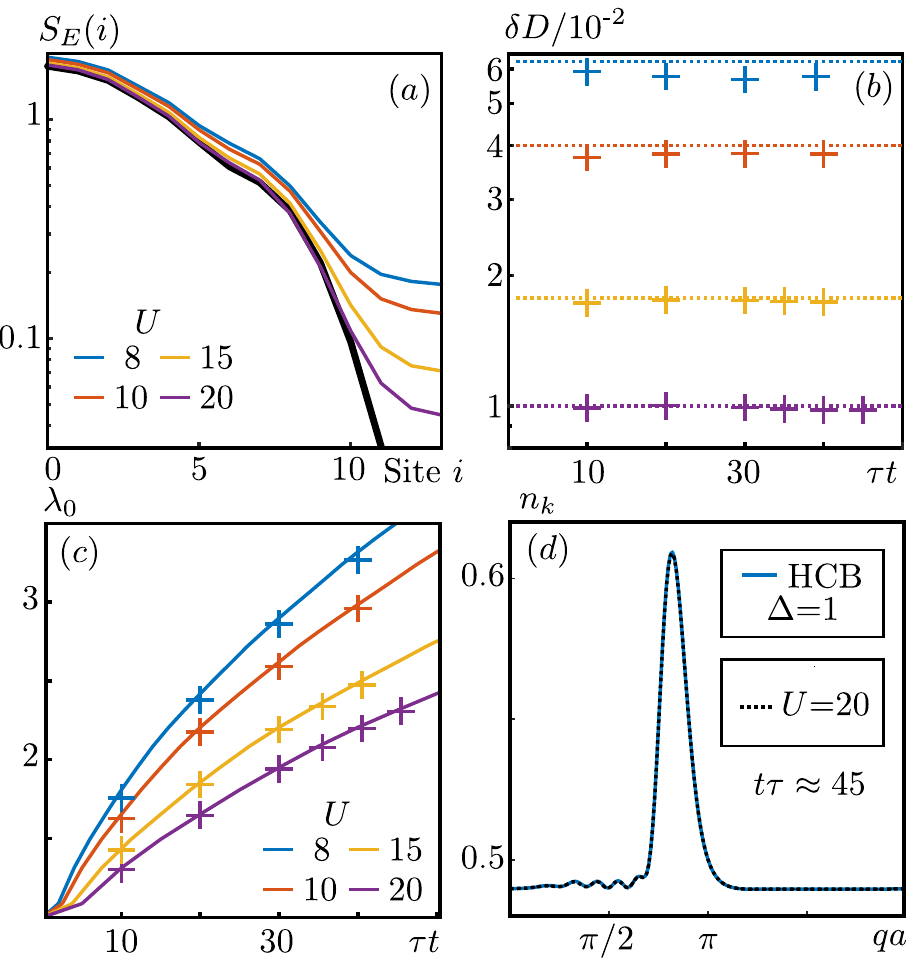}
	\caption{(a) The entanglement entropy across the system for the effective model with $J= 4t^2/U$ (black line) at $J\tau = 10$ compared with the undriven Hubbard model for a selection of $U$'s. (b) The loss in the number of doublons $\delta D$ in the undriven Hubbard model the same set of $U$'s. (c) The value of the LNO occupation of the undriven Hubbard model $(+)$ with time for the same $U$'s compared to the effective model (solid lines). (d) The momentum distribution of the entire system for the undriven Fermi-Hubbard model and the effective model. These calculations used $L = 200$ sites, $t\delta\tau = 0.005$ and an MPS bond dimension $\chi = 2000$. The effective model calculated on $L=200$ sites with $N=100$ doublons and MPS bond dimension $\chi = 1200$.}
	\label{FIG:Undriven_Doublon_Num_and_LNO}
\end{figure}

Despite having a finite $U/t$ the decay of doublons quickly saturates with time. This is confirmed by computing the deviation in the total number of doublon number $\delta D(\tau) = \av{\hat{D}(0)} - \av{\hat{D}(\tau)}$ with $\hat{D} = \sum_i \hat{n}_{d,i}$ in the time-evolved state. Second order time-dependent perturbation theory predicts this observable will behave as
\begin{align}
 \delta D(\tau) = 8\left(\frac{t}{U}\right)^2\sin^2\left(\frac{1}{2}U\tau\right). \label{EQN_td_pert}
\end{align}
In \fir{FIG:Undriven_Doublon_Num_and_LNO}(b) we plot $\delta D(\tau)$ for several $U/t$'s showing that it saturates to a constant given by the time-average of \eqr{EQN_td_pert} and is thus suppressed as $(t/U)^2$. 

Given the bounded fraction of fermions generated by finite $U/t$ we next examined the key characteristics of quasicondensation in the full Hubbard model. In \fir{FIG:Undriven_Doublon_Num_and_LNO}(c) the LNO occupation of the doublon OPDM computed from the full Hubbard model is compared to the effective model with the appropriate superexchange $J = 2t^2/U$. We find good agreement for the LNO growth with time, even for very moderate interaction strengths $U/t = 8$. In \fir{FIG:Undriven_Doublon_Num_and_LNO}(d) we verify that a peak in the momentum distribution of the melt-lobe continues to manifest close to $q_{\rm c}a = \pi$ in agreement with the effective model at $\Delta = 1$. These results together confirm that the quasicondensation seen in the effective model is extremely robust to the presence of a small amount of initial doublon disassociation. The reason for this is that the fermions rapidly expand beyond the melt lobe region leaving it essentially undisturbed. Such a distillation of doublons and fermions was recently demonstrated experimentally \cite{Scherg2018}.  

\begin{figure*}[t]
	\centering
	\includegraphics{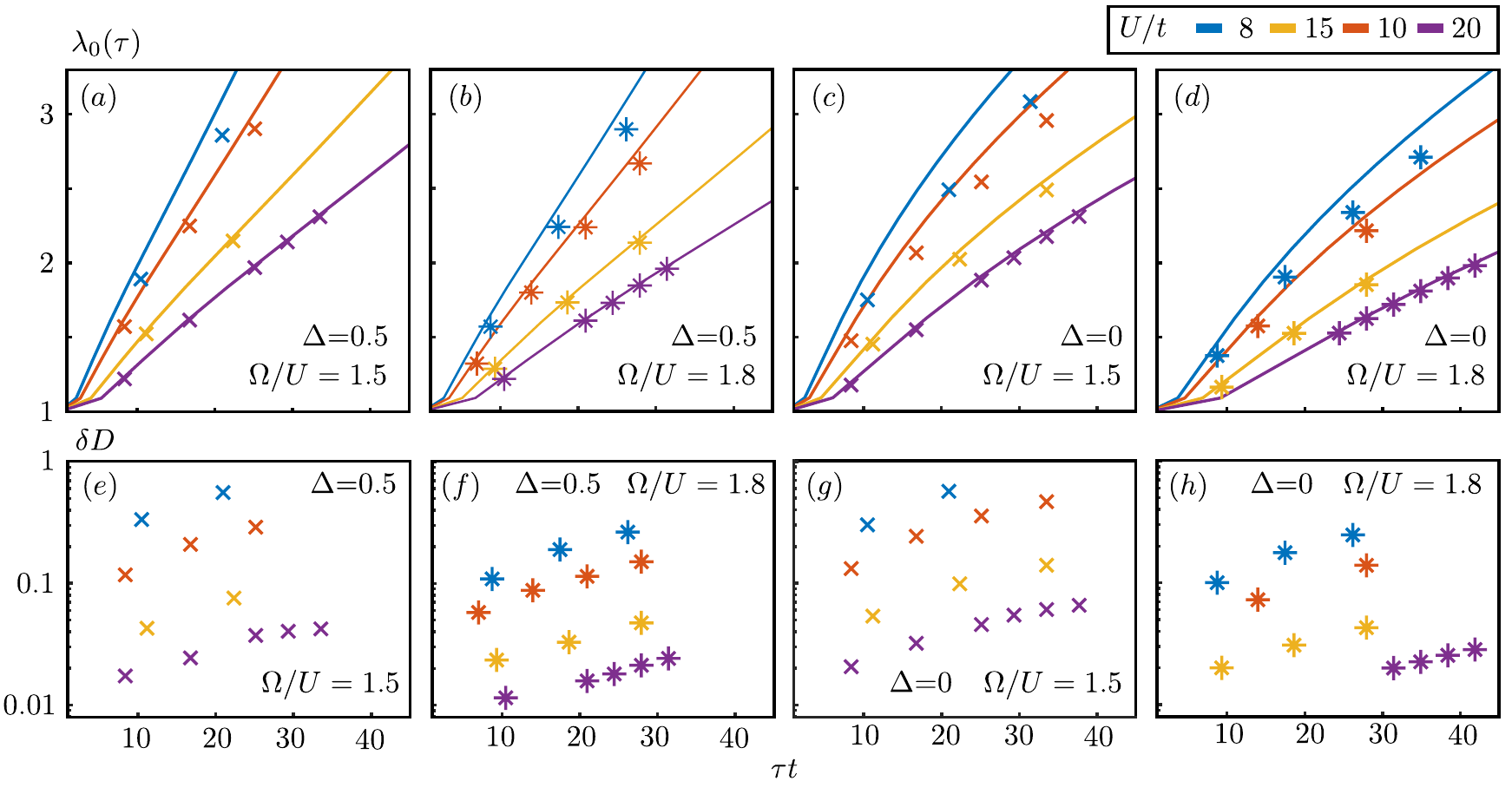}
	\caption{(a)-(d) A comparison between the LNO amplitudes during the melting of an effective model of hard core bosons (solid lines) and the full driven Hubbard model (points) for different interaction strengths, frequencies and $\Delta$'s of the effective model. The driven system is driven with an amplitude large enough that the anisotropy $\Delta(\nu,\Omega)$ in \eqr{Eqn:DrivenAnisotropy} is $\Delta = 0.5$ in (a),(b) and $\Delta = 0$ in (c),(d). Two frequencies have been used, $\Omega = 1.5U$ in (a),(c) and $\Omega = 1.8U$ in (b),(d).  Bottom row (e)-(h), the loss in the number of doublons $\delta D(\tau) = D(0) - D(\tau)$ where $D = \left< \sum_i n_{i \uparrow}n_{i \downarrow}  \right>$ for the same parameter sets used in the figures above, on a log-linear scale. All system sizes for the Fermi Hubbard model are $L=200$, with internal dimension $\chi = 2000$ and time step $t \delta \tau =2\pi/(50 \Omega)$. The comparison to the effective model used calculation in \fir{FIG:OPDM_3D}}
	\label{FIG:DRIVEN_LNO_DOUBLON_MOM}
\end{figure*}

\begin{figure}[t]
	\centering
	\includegraphics[scale=0.9]{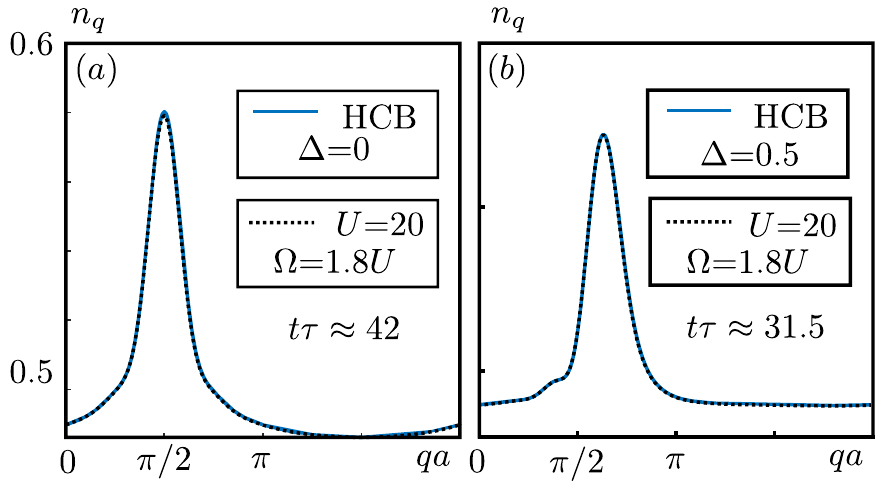}
	\caption{The momentum distributions of doublons in the full driven Fermi-Hubbard model (black dotted line) for two driving strengths corresponding to (a) $\Delta =0$ and (b) $\Delta = 0.5$ evaluated at the driving period commensurate time $\tau$ specified. In both cases this and the effective model (blue solid line) at the same time with a superexchange \eqr{Eqn:DrivenSuperExchange} and anisotropy \eqr{Eqn:DrivenAnisotropy} is also shown. Calculation details of the Fermi Hubbard model are identical to \fir{FIG:DRIVEN_LNO_DOUBLON_MOM}. Effective model calculated on $L=200$ sites with $N=100$ particles and MPS bond dimension $\chi = 1200$.}
	\label{FIG:DRIVEN_SF_D0_Compare}
\end{figure}

\subsection{Driven case}
\label{sec:driven_hubbard}
We now address how well the effective model quench describes the full Fermi-Hubbard model dynamics with an abrupt application of finite frequency driving. Since the Floquet effective model is formally only a stroboscopic description of the driven Fermi-Hubbard model we compare the models in this section at times $\tau$ that are integer multiples of the driving period $T$. This ensures that both the kick operators $K(\tau)$ and transformation $R(\tau)$ from the lab to the rotating frame are equal to identity. We focus on two representative cases with the driving amplitude $\nu$ tuned to a moderate strength, so $\Delta(\nu,\Omega) = 0.5$, and also stronger so $\Delta(\nu,\Omega) = 0$. In \fir{FIG:DRIVEN_LNO_DOUBLON_MOM}(a)-(d) we plot the LNO occupation in time of the driven Hubbard model for these two $\Delta(\nu,\Omega)$'s for range of $U$'s and two different driving frequencies $\Omega/U = 1.5$ and $\Omega/U = 1.8$. We observe good agreement with the growth predicted by the effective model over the times examines, despite the finite interactions and driving frequency. As expected the agreement improves when increasing $U/t$ and/or $\Omega/U$. In \fir{FIG:DRIVEN_LNO_DOUBLON_MOM}(e)-(h) we plot the corresponding $\delta D$. For the driven Hubbard model we observe an order of magnitude increase in the doublon loss compared to the undriven case in \fir{FIG:Undriven_Doublon_Num_and_LNO}(a). Furthermore the doublon loss does not saturate and instead displays a roughly exponential increase in time with a rate constant that is suppressed with increasing $\Omega$. This behaviour is a remnant of Floquet heating. The doublon losses also explain the more noticeable lag in the LNO population growth for the lowest values of $U/t$, e.g. seen in \fir{FIG:DRIVEN_LNO_DOUBLON_MOM}(a) and (c). 

Given the exponential doublon losses we estimate that in the worse case considered, namely $U/t = 8$, $\Omega/U = 1.5$ and $\Delta(\nu,\Omega) = 0$ in \fir{FIG:DRIVEN_LNO_DOUBLON_MOM}(g), that the quasicondensate predicted by the effective model will be completely depleted by $t\tau \sim 50$. For more favourable parameters, such as $U/t = 20$ and $\Omega/U = 1.8$ in \fir{FIG:DRIVEN_LNO_DOUBLON_MOM}(h), this time is significantly extended to $t\tau \sim 200$. In this case, examining $t\tau \approx 42$ below this time where doublon loss is negligible, \fir{FIG:DRIVEN_SF_D0_Compare}(a) shows that the doublon momentum distribution displays a sharp peak positioned at $q_{\rm c}a = \pi/2$. Remarkably this is very close what is expected from the free expansion of non-interacting hard core bosons in the effective model (also shown), yet is realised here in a driven, strongly interacting fermionic Hubbard model with a moderate drive frequency $\Omega \gtrsim U$. In \fir{FIG:DRIVEN_SF_D0_Compare}(b) at weaker driving where $\Delta(\nu,\Omega) = 0.5$ we find a peak at $q_{\rm c}a > \pi/2$. Together with the zero driving peak in \fir{FIG:Undriven_Doublon_Num_and_LNO}(d) approaching $q_{\rm c}a = \pi$, we see that a controllable range of quasicondensing momenta are indeed accessible in the driven Hubbard model.

\subsection{Cold-atom implementation}
\label{sec:cold_atom_setup}

\begin{figure}[t]
	\centering
	\includegraphics{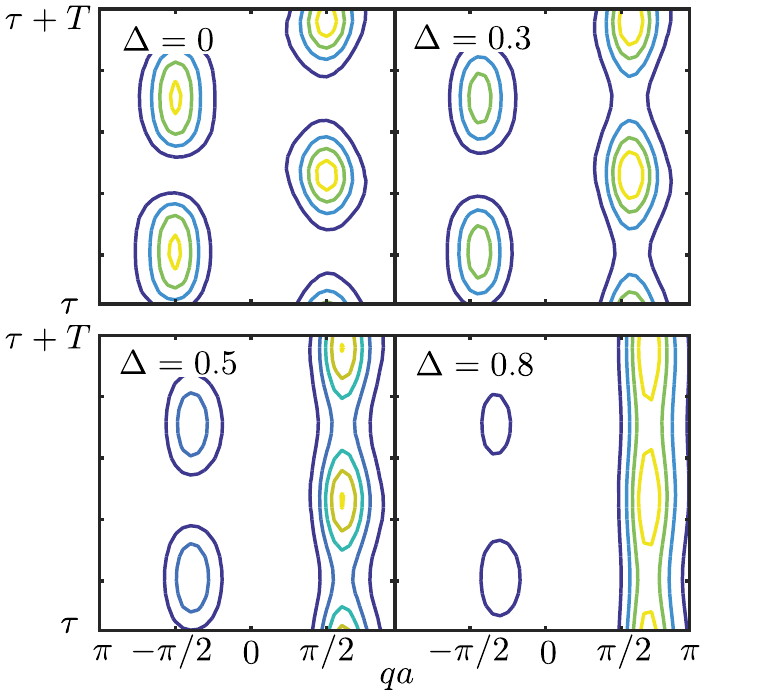}
	\caption{Contour maps of the evolution of the momentum distribution over a single driving period. The system was evolved to $\tau = 80T \approx 16.7/t$ with the driven Hubbard model using $U/t=20$, $\Omega/t = 30$ and four different driving strengths corresponding to $\Delta(\nu,\Omega) = 0,0.3,0.5$ and $0.8$. System size $L=40$ with $N=20$ of each species and MPS bond dimension $\chi = 2000$.  }
	\label{FIG:INTERPERIOD_SF}
\end{figure}

\begin{figure}[t]
	\centering
	\includegraphics{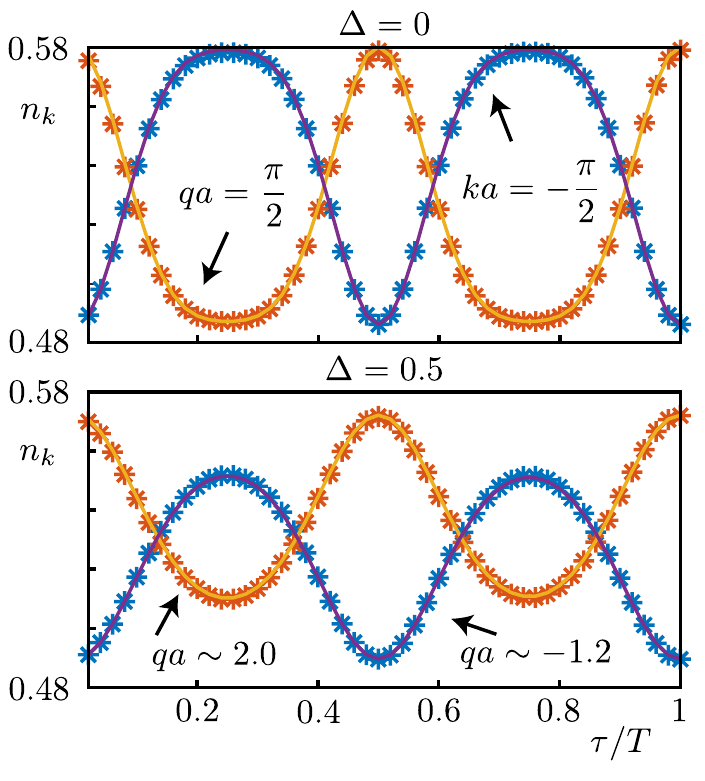}
	\caption{ The height of the momentum distribution taken at two given momenta which are separated by $\delta q a = \pi$. The height is plotted as a function of time over a single driving period. Solid lines are the result of the effective model transformed back into the lab frame (the transformation $R^\dagger(\tau)$ of \eqr{EQN:Rotating_Frame_Transform} has been applied), while $*$'s are obtained from the momentum distribution of the driven Fermi-Hubbard model using the entire $L=200$ site system over one driving period $T = 2\pi/\Omega$. Calculational details idential to \fir{FIG:DRIVEN_LNO_DOUBLON_MOM}}
	\label{FIG:INTERPERIOD_SF_2}
\end{figure}

The driven dynamical quasicondensation outlined is realizable in a standard optical lattice experiment with feasible lattice parameters \cite{Kohl2005,TARRUELL2018,Scherg2018,Esslinger2010}. To illustrate this we consider the well studied case of fermionic K$^{40}$ in an undriven 3D optical lattice potential. One dimensional systems are realized by choosing anisotropic depths $V_x = 8E_R$ along the axis of the chains and $V_y=V_z=33E_R$ for the transverse confinement, where $E_R=\hbar^2 k^2_L / 2 m_a$ is the recoil energy with $\lambda_L = 2 \pi / k_L = 532$nm being the laser wavelength and $m_a$ is the atomic mass. For this system the nearest-neighbour hopping amplitude of the Hubbard model is $t \sim 0.54$kHz, with small next-nearest-neighbour $t' \sim t/50$. Given typical fermionic cold-atoms systems can remain coherent for up to 1s this in principle allows for a total experiment time of $t\tau \sim 500$. This is consistent with a central domain of order 100 hundred sites surrounded by similarly sized empty regions. The band-insulating initial state can be generated using additional magnetic trapping and tuning the system to have attractive interactions via a Feshbach resonance.  

The implementation of the alternating potential driving term included in the Hamiltonian \eqr{EQN:Ham_Drv_Part} is slightly non-standard, but is nonetheless realisable with state of the art optical lattice experiments. Conventionally an optical lattice is driven by shaking the entire trapping potential by a length $\delta(\tau) = \delta_0 \cos(\Omega \tau)$ via kHz piezoelectric modulation of lattice laser's phase. Shaking along the $x$-axis then induces an effective linear potential in the non-inertial reference frame $V(x) = - \Omega^2 \delta_0 x$. To create an alternating driving potential we propose a scheme based on a zig-zag chain geometry \cite{Zhang2015,Anisimovas2016} where odd and even sites are laterally displaced. Shaking perpendicular to the chain but parallel to the zig-zag then induces an effective modulated potential difference between odd and even sites. More details and parameter estimations for this setup are discussed in Appendix \ref{SEC:ZigZagLattice}.

Neither time-of-flight nor in-situ measurements of cold-atom will implement perfect stroboscopic sampling of the driven system. As such we now examine how the momentum distribution of the driven Fermi-Hubbard model deviates from the effective model at times inside a single driving period. Specifically, we time-evolved the full driven Hubbard model in the lab frame given in \eqr{EQN:Driven_Hubbard_Model} with $U/t = 20$ until $\tau = 80T \approx 16.7/t$ and then frequently measured the momentum distribution over the next driving period at small increments $\Delta t \approx 0.05T$. The contour plots of the resulting distributions for various driving strengths are displayed in \fir{FIG:INTERPERIOD_SF}. The contour plots begin at a commensurate time $\tau = 80T$ so as expected a peak centred at the $q_c$ of the corresponding effective model's $\Delta$ is seen. As time progresses within the drive period the momentum distribution oscillates at a frequency $2\Omega$ by exchanging weight from the primary peak to another secondary peak shifted by a momentum $\pi/a$. The relative amplitude of the secondary peak depends on the driving strength, starting at zero for the undriven $\Delta = 1$ case and reaching unity for the strongly driven $\Delta = 0$ case.

We confirmed that the transformation from the rotating frame $R(\tau)$ (given by \eqr{EQN:Rotating_Frame_Transform} in Appendix \ref{sec:app_floquet}) is the origin of this oscillatory behaviour. In \fir{FIG:INTERPERIOD_SF_2} the amplitude of the primary and secondary peaks for strong and moderate driving strengths from the full driven Hubbard model's momentum distribution  are directly compared to those of the effective model's after transforming back to the lab frame. The excellent agreement between the two models indicates that the kick operators have a negligible effect for this observable. Furthermore, the appearance of two distinguishable momentum peaks for measurements at general times is a desirable experimental signature of the alternating driving scheme proposed here since it is robust to time-averaging.  

\section{Conclusions}
\label{sec:conclusion}
We have demonstrated that dynamical quasicondensation at a controllable finite momentum can emerge from a band-insulating domain initial state whose expansion is governed by a one-dimensional driven Hubbard model in the strongly-interacting regime. To establish this we first examined an Floquet effective model. We showed how the known dynamical quasicondensation of non-interacting $\Delta = 0$ hard-core bosons at $q_{\rm c}a = \pi/2$ not only persists in the interacting systems with doublon-hole repulsion $0< \Delta \leq 1$, but does so with a stronger LNO growth and at a momentum that shifts towards $q_{\rm c}a = \pi$. However, our numerical calculations were not fully conclusive on the nature of this quasicondensation owing to the slowing down of the doublon propagation with increasing $\Delta$ which limited the system sizes accessible. We found significant differences between the melt state and a boosted ground state for the times examined, in contrast to hydrodynamical calculations for soft domain walls. We cannot rule out that these differences are transient and that such a correspondence may yet emerge at much longer times when the initial hard domain has been softened by the expansion. Next, we went beyond the effective model and simulated the full driven Hubbard model. We demonstrated good agreement with the predictions of the effective model for key observables over short times, even at very moderate finite interactions and driving frequencies. The presence of Floquet heating induced doublon loss, not captured by the effective model, was shown to be present but does not preclude quasicondensation for the times examined.

Given that all the ingredients of the setup proposed have been implemented in current state-of-the-art cold-atom setups the effects outlined are in principle within reach of experimental observation~\cite{Bloch2008}. Indeed a cold-atom quantum simulator might be the most definitive means of answering the open question as to whether dynamical quasicondensation of doublons also occurs in higher spatial dimensions. Moreover, our quench setup presents a potentially fast and robust scheme for creating a much sought-after $\eta$-quasicondensate with cold-atoms~\cite{Rosch2008,Kantian2010}.  

As an outlook it is interesting to speculate whether dynamical quasicondensation can be relevant to experiments on driven solid-state systems. Several features make this plausible. First, owing to the beam spot-size and limited penetration depth of all pump-probe experiments so far, only a small excitation volume of sample is driven \cite{Giannetti2016}. Consequently the induced excited state is necessarily inhomogeneous and it is an open question whether the subsequent expansion dynamics of charge-carriers out of this excitation volume into the rest of the material plays a crucial role in the non-equilibrium superconducting coherence observed \cite{Kaiser2014b,Mitrano2016}. Second, a sharper domain, more similar to that considered here, can be engineered in a solid by heterostructuring. For example, the time-resolved scrambling of magnetic order in a thin-film due to the expansion of interfacial shockwaves of mobile carriers from a substrate has been observed \cite{Forst2015}. The possibility of 
observing instead the dynamical ultra-fast emergence of order from a shockwave is intriguing.

\section{Acknowledgements}
MWC thanks the University of Bath for URSA funding. SRC gratefully acknowledges support from the UK’s Engineering and Physical Sciences Research Council (EPSRC) under grant No. EP/P025110/1.

\appendix

\section{Floquet expansion} \label{sec:app_floquet}
Floquet's theorem dictates that any Hamiltonian with periodic time dependence will have solutions to the time dependent Schr\"{o}dinger equation $H(\tau) \Psi(\tau) = \mi \partial_{\tau}\Psi(\tau)$ of the form $\ket{\psi_{n}(\tau)} = \exp(-\mi \varepsilon_n \tau)\ket{\phi_n(\tau)}$ where $\varepsilon_n$ is the quasi-energy associated with the $T$-periodic Floquet state $\ket{\phi_n(\tau + T)} = \ket{\phi_n(\tau)}$. In the basis $\{\ket{\phi_n(\tau)}\}$ the time dependent Schr{\"o}dinger equation is recast as
	\begin{align}
	    \label{EQN:FTDSE}
			F \ket{\phi_n(\tau)} = \left[H(\tau) - \mi \partial_\tau \right] \ket{\phi_n(\tau)} = \varepsilon_n \ket{\phi_n(\tau)},
	\end{align}
where $F = H(\tau) - \mi \partial_\tau$ is the Floquet quasienergy operator.

The task of finding the quasi-energies $\varepsilon_n$ and the Floquet modes $\ket{\phi_n(\tau)}$ can be achieved by diagonalising $F$ over an extended Hilbert space $\mathcal{F}$. Specifically, $\mathcal{F} = \mathcal{H} \otimes \mathcal{T}$ is the tensor product space of the original Hilbert space $\mathcal{H}$ and the space $\mathcal{T}$ of functions $f(\tau) = f(\tau+T)$ with periodicity $T = \frac{2 \pi}{\Omega}$. We denote states in $\mathcal{T}$ with a double angled bracket $| f \rrangle$ and use a scalar product
\begin{align}
    \llangle f,g \rrangle = \frac{1}{T} \int_0^T d \tau f^*(\tau) g(\tau).
\end{align}
In this derivation it is convenient to choose the Fourier basis of time periodic functions $|n\rrangle = \exp(\mi n \Omega \tau)$ that are eigenstates of $\partial_\tau$ with $\partial_\tau |n\rrangle = \mi n \Omega |n\rrangle$ obeying $\llangle n, m \rrangle = \delta_{n,m}$. The index $ n \in \mathbbm{Z}$ is often called the ``photon number'' of a Floquet sector. The matrix elements of the quasi-energy operator $F$ using this basis for $\mathcal{T}$ and any basis for $\mathcal{H}$ are then 
\begin{align}
\llangle m | \left< v \right| F  \left| u \right> | n \rrangle  =  \llangle m | \left< v \right|  H(\tau)  \left| u \right> | n \rrangle   + n \Omega \delta_{n,m}.
\end{align}
Since $\llangle m |\left< v \right|  H(\tau)  \left| u \right> | n \rrangle$ only depends on the difference $(n-m)$, thus the quasi-energy operator $F$ contains replicas of $H$ for each photon number $n$, each shifted in energy by $n\Omega$.  Despite the infinite duplication of the many-body Hilbert space, for very high-frequency driving, where $\Omega$ is larger than all matrix elements, standard degenerate perturbation theory can be used to derive an approximate description of a single-band due to corrections from the neighbouring bands. 

Here we describe a strong coupling expansion onto both a single Floquet sector and onto a specific band of Hubbard eigenstates with a given doublon number. In the lab frame, the coupling between the Floquet sectors is only non zero for $n-m =\pm1$, and the coupling strength is comparable to $\Omega$. This is an unsuitable starting point for perturbation theory. However, the coupling can be reduced by transforming into a frame rotating with the driving, where the time dependence becomes a Peierls phase with periodic time dependence. This is achieved with the unitary 
\begin{align}
\label{EQN:Rotating_Frame_Transform}
    R(\tau) =& \exp{ \left[ \mi \int_0^\tau d\tau' H_{\text{drv}}(\tau') \right] } \\
	 =& \exp{ \left[  \mi \frac{A}{2 \Omega}   \sin(\Omega \tau) \sum_{i \sigma} \left(-1 \right)^i n_{i \sigma}\right] },
\end{align}
which transforms the Hamiltonian into the rotating frame as 
\begin{align}
\bar{H}(\tau) = \mi \dot{R}(\tau)R^\dagger(\tau) + R(\tau)H(\tau)R^\dagger(\tau).	
\end{align}
The first term removes the driving potential $H_{\text{drv}}(\tau)$, and while the second term puts a time-dependent momentum shift onto the kinetic energy term as
\begin{align}
	\begin{split}
	R(\tau) & H_{\text{kin}}R^\dagger(\tau) \\
	=& -t \sum_{i \sigma}  \exp{\left[\mi \frac{A}{ \Omega}  \sin(\Omega \tau)  \left(-1 \right)^i \right]} c^\dagger_{i \sigma} c_{i + 1 \sigma}   + \text{h.c.}\\
	=& -t \sum_{i \sigma m } \mathcal{J}_m(\nu) \exp{ \left[\mi m \Omega \tau(-1)^\di ) \right]} c^\dagger_{i \sigma} c_{i + 1 \sigma} + \text{h.c.},
	\end{split}
\end{align}
where $\mathcal{J}_m(\cdot)$ is the Bessel function of the first kind and $\nu=A/\Omega$ is the dimensionless driving strength.   

A perturbative expansion around the relevant set of eigenstates requires two projectors. First, $\mathbbm{P}_n$ projects onto the eigenstates with $D$ double occupations
\begin{align}
    \mathbbm{P}_D =& \frac{(-1)^D \partial^D_\alpha}{D!}  \left[ \prod_i \left.\left( 1- \alpha n_{i \uparrow} n_{i \downarrow} \right) \right] \right|_{\alpha = 1},
\end{align}
and $\mathbbm{M}_n$ projects onto the $n$ photon sector
\begin{align}
    \mathbbm{M}_n =& | n \rrangle \llangle n|.
\end{align}
We define the small parameter $\lambda = t/U$ and let the frequency $\Omega$ be of comparable magnitude to the interaction strength $U \sim \Omega \gg t$. We then define the dimensionless Floquet quasi-energy operator in the rotating frame as 
\begin{align}
    \bar{F}' = \bar{F}/ U = \bar{F}'_0 + \lambda \bar{F}_1',
\end{align}
where $\bar{F}'_0 = \sum_{D,n} \mathbbm{P}_D \otimes \mathbbm{M}_{n}\, E_{D,n}$ is the operator which counts doublon number and photon index, with a highly degenerate set of eigenvalues $E_{D,n} = D + \left(\Omega/U\right)n$. The perturbation $\bar{F}'_1$ contains the kinetic coupling of eigenstates of different doublon and photon as
\begin{align}
    \bar{F}'_1 = - \sum_{i \sigma m n } \mathcal{J}_m(\nu) c^\dagger_{i \sigma} c_{i + 1 \sigma} \otimes |  n+m(-1)^\di \rrangle \llangle n | + \text{h.c.}.
\end{align}

We denote the perturbative expansion of the quasi-energy operator $\bar{F}'$ about the set of eigenstates with exactly $D$ doublons and $n$ photons as $\bar{F}'_{D,n}$. For the relevant case of the maximum doublon number $D=N$ and the DC Floquet sector $n=0$ we have
\begin{align}
	\begin{split}
	    \bar{F}'_{N,0} = & E_{N,0} + \lambda \left( \mathbbm{P}_N \otimes \mathbbm{M}_{0} \right) \bar{F}'_{1} \left(\mathbbm{P}_N \otimes \mathbbm{M}_{0} \right) \\
	    + & ~ \lambda^2 \sum_{a \neq 0, b \in \mathbbm{Z}} \frac{\left(\mathbbm{P}_N \otimes \mathbbm{M}_{0}\right) \bar{F}'_1 \left(\mathbbm{P}_a \otimes \mathbbm{M}_{b}\right) \bar{F}'_1 \left(\mathbbm{P}_N \otimes \mathbbm{M}_{0}\right)}{E_{N,0} - E_{a,b}}.
	\end{split}
\end{align}
To second order the only contributions in this effective model are from consecutive hops which break up and reform a doublon. This reformation can occur on the same site the doublon started on, or on a neighbouring site. Since the sign of the Floquet transitions depends on the direction of the hopping, the processes involving two hops in a single direction will pick up different amplitudes to processes which are two hops in opposite directions. 

To compute these amplitudes we define two operators. The first is
\begin{align}
h^\dagger_{ij\sigma} &=  n_{i \bar{\sigma}} c^\dagger_{i\sigma} c_{j \sigma}  (1-n_{j\bar{\sigma}}),
\end{align}
which creates a doublon on site $i$ when a fermion of spin $\sigma$ occupied site $j$ and a fermion of spin $\bar{\sigma} = -\sigma$ occupied site $i$. The second is
\begin{align}
g_{ij\sigma} &= (1-n_{i\bar{\sigma}}) c^\dagger_{i\sigma} c_{j \sigma} (1-n_{j\bar{\sigma}}) +n_{i\bar{\sigma}} c^\dagger_{i\sigma} c_{j \sigma} n_{j\bar{\sigma}},
\end{align}
which describes the hopping of singly occupied sites over either a doublon or a hole. Using these operators the term $c^\dagger_{i\sigma} c_{j \sigma}$ describing hopping of a spin $\sigma$ fermion from $j$ to $i$ is decomposed into 
\begin{align}
	c^\dagger_{i\sigma} c_{j \sigma}  = h^\dagger_{ij\sigma} + h_{j i\sigma} +  g_{ij\sigma}. 
\end{align}
The energy change to second order is $E_{N,0} - E_{N-1,b} = U - b \Omega$, and the numerator in the expansion of $\bar{F}'_{N,0}$ is made of two parts. We automatically drop the $g$ and $h^\dagger$ contributions from $\bar{F}'_1 \left(\mathbbm{P}_N \otimes \mathbbm{M}_{0}\right)$, since there is no single fermion hop which preserves doublon number and no free fermions available to create a doublon. We similarily drop all $h$ and $g$ type from $\left(\mathbbm{P}_N \otimes \mathbbm{M}_{0}\right)\bar{F}'_1 $ for the same reason. Finally we remove the projectors from the middle of the expression since we are automatically guaranteed to step down by one doublon. The expression for $\bar{F}_{N,0}'$ then becomes
\begin{widetext}
\begin{align}
    \bar{F}'_{N,0} = &  \lambda^2 \sum_{ b i j \sigma \tau nm n' m'} \left(\mathbbm{P}_N \otimes \mathbbm{M}_{0}\right)  \mathcal{J}_m(\nu)\Big( \left[  h^\dagger_{j+1,j,\sigma} |  n+m(-1)^j \rrangle \llangle n | \right] + \left[ h^\dagger_{j,j+1,\sigma} |  n \rrangle \llangle n+m(-1)^j | \right] \Big) \nonumber \\
    & \mathcal{J}_{m'}(\nu) \Big(  \left[  h_{i+1,i,\tau} |  n'+m'(-1)^\di \rrangle \llangle n | \right] + \left[ h_{i,i+1,\tau} |  n' \rrangle \llangle n'+m'(-1)^\di | \right] \Big) \left(\mathbbm{P}_N \otimes \mathbbm{M}_{0}\right) / \left( {E_{N,0} - E_{a,b}} \right). 
\end{align}
Multiplying out the two hopping processes gives
\begin{align}
    \bar{F}'_{N,0} = &\lambda^2 \sum_{ b i j \sigma \tau nm n' m'} \mathcal{J}_m(\nu) \mathcal{J}_{m'}(\nu)\left(\mathbbm{P}_N \otimes \mathbbm{M}_{0}\right) \nonumber \\
    &  \Big( \left[  h^\dagger_{j+1,j,\sigma}h_{i+1,i,\tau} |  n+m(-1)^j \rrangle \llangle n  |  n'+m'(-1)^\di \rrangle \llangle n' | \right] + \left[ h^\dagger_{j,j+1,\sigma}h_{i+1,i,\tau} |  n \rrangle \llangle n+m(-1)^j |   n'+m'(-1)^\di \rrangle \llangle n' | \right]  \nonumber \\
    &+ \left[h^\dagger_{j+1,j,\sigma}h_{i,i+1,\tau} |  n+m(-1)^j \rrangle \llangle n |   n' \rrangle \llangle n'+m'(-1)^\di |  \right] + \left[ h^\dagger_{j,j+1,\sigma}h_{i,i+1,\tau} |  n \rrangle \llangle n+m(-1)^j |   n' \rrangle \llangle n'+m'(-1)^\di | \right] \nonumber \Big) \nonumber \\
    &\left(\mathbbm{P}_N \otimes \mathbbm{M}_{0}\right) / \left( {E_{N,0} - E_{a,b}} \right). 
\end{align}
The sum over lattice site $j$ can be removed by noting that unless the doublon is created by $h^\dagger$ on one of the same two sites as $h$, we will change doulbon number. Therefore the only allowed combinations are of the form $h^\dagger_{i,j} h_{i,j}$ or $h^\dagger_{j,i} h_{i,j}$. We also replace the inner products in $\mathcal{T}$ with the constraints $\llangle n | m \rrangle = \delta_{n,m}$ to give
\begin{align}
    \bar{F}'_{N,0} = &\lambda^2 \sum_{ b i j \sigma \tau nm n' m'} \mathcal{J}_m(\nu) \mathcal{J}_{m'}(\nu)\left(\mathbbm{P}_N \otimes \mathbbm{M}_{0}\right) \nonumber \\
    & \Big( \left[\delta_{i,j}\delta_{n,n'+m'(-1)^i} \delta_{n',0}\delta_{n+m(-1)^i,0} h^\dagger_{j+1,j,\sigma}h_{i+1,i,\tau} \right] + \left[\delta_{i,j}\delta_{n,0} \delta_{n+m(-1)^j,n'+m'(-1)^\di} \delta_{n',0}h^\dagger_{j,j+1,\sigma}h_{i+1,i,\tau}\right] \nonumber \\
    &+ \left[ \delta_{i,j}\delta_{n+m(-1)^j,0}\delta_{n,n'}\delta_{n'+m'(-1)^\di,0} h^\dagger_{j+1,j,\sigma}h_{i,i+1,\tau} \right] + \left[ \delta_{i,j} \delta_{n,0}\delta_{n+m(-1)^j,n'}\delta_{n'+m'(-1)^\di,0} h^\dagger_{j,j+1,\sigma}h_{i,i+1,\tau} \right] \Big) \nonumber \\
    &\left(\mathbbm{P}_N \otimes \mathbbm{M}_{0}\right) / \left( {U - b\Omega} \right).
\end{align}
\end{widetext}
We omit the left and right projectors $\mathbbm{P}_N \otimes \mathbbm{M}_{0}$ from here on since their presence is implied. The three remaining constraints on the four indices denoting the Floquet sectors leaves one independent photon index to sum over as 
\begin{align}
    \begin{split}
         = \lambda^2 \sum_{  i \sigma  m} &\frac{\mathcal{J}_m(\nu) \mathcal{J}_{-m}(\nu)}{ U + m\Omega} \left[ h^\dagger_{i+1,i,\sigma} h_{i+1,i,\sigma} + h^\dagger_{i,i+1,\sigma}h_{i,i+1,-\sigma} \right] \\
        + &\frac{\mathcal{J}_m(\nu) \mathcal{J}_{m}(\nu)}{ U + m\Omega} \left[ h^\dagger_{i,i+1,\sigma} h_{i+1,i,-\sigma}   
        + h^\dagger_{i+1,i,\sigma}h_{i,i+1,\sigma} \right].
    \end{split}
\end{align}
The two parts to the Hamiltonian can be interpreted as nearest-neighbour repulsion of a doublon and a hole, 
\begin{align}
	 h^\dagger_{i+1,i,\sigma} &h_{i+1,i,\sigma} =n_{i+1,\sigma}n_{i+1,-\sigma} (1-n_{i,\sigma})(1-n_{i,-\sigma}),
\end{align}
and a nearest-neighbour hopping of doublons in one direction and holes in the opposite direction, 
\begin{align}
	 h^\dagger_{i,i+1,\sigma} & h_{i+1,i,\sigma}  = c^\dagger_{i,\sigma}c^\dagger_{i,-\sigma}c_{i+1,-\sigma}c_{i+1,\sigma}. 
\end{align}
The two amplitudes given in \eqr{Eqn:DrivenAnisotropy} and \eqr{Eqn:DrivenSuperExchange} are recovered since $\mathcal{J}_{-m}(\nu) = (-1)^m\mathcal{J}_{m}(\nu)$. 

\section{Implementing a zig-zag lattice}

\begin{figure}[t]
	\includegraphics{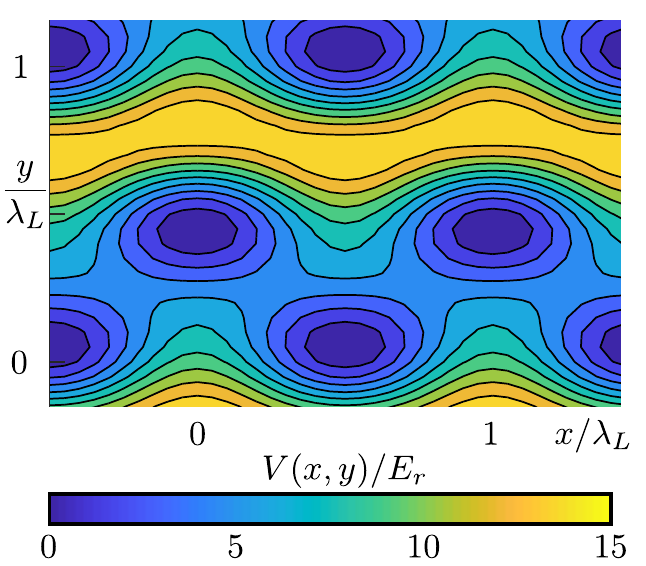}
\caption{The potential which creates the zig-zag optical lattice plotted here for $V_x = -10E_R$, $V_y = V_x$, $V_{\bar{x}} = 2 V_x$ and $V_{\text{sq}} = 0$.} \label{FIG:ZigZagPot}
\end{figure}

\label{SEC:ZigZagLattice}
One possible way of creating a driving term of the form \eqr{EQN:Ham_Drv_Part} is using a one dimensional chain with a lateral displacement of nearest-neighbour sites, i.e. a zig-zag type geometry. Proposals for creating a zig-zag type lattice (aswell as rhombic, sawtooth, etc.) are detailed in Ref.~\cite{Zhang2015}. With a specific choices of lattice laser phases they derive a potential of the form
\begin{align}
V&(x,y) =  V_x \cos^2(k_L  x) + \sqrt{V_x V_y} \cos(k_L x) \cos(k_L y)  \nonumber\\
 & + V_{\bar{x}} \cos^2(k_L  x/2 - \pi/4) \nonumber \\ 
&+  V_{\text{sq}} \left(\cos^2(k_L  x/2) + \cos^2(k_L  y/2)\right),
\end{align}
which is capable of creating a ``zig-zag'' type lattice under certain parameter choices. A suitable example is as follows, we set $V_{\text{sq}}=0 $ to ensure the onsite energy of each Wannier orbital is equal, while $V_x = -10E_R$, $V_y = V_x$, $V_{\bar{x}} = 2 V_x$. Here $E_R$ is the recoil energy, which for potassium-40 is $E_R \approx 17$kHz. An additional deep confinement potential in the $z$ direction of $-20E_R \cos^2(k_L z)$ is used to achieve 1D tubes. The resulting potential is shown in \fir{FIG:ZigZagPot}.

For this choice of parameters we have computed the dominant Hubbard Hamiltonian matrix elements using the Wannier MATLAB package as detailed in Ref.~\cite{Walters2013}. We find that the hopping amplitudes displayed in \fir{FIG:ZigZagHop} are $t = 0.03 E_R \approx 0.6$kHz for nearest-neighbour hopping between A and B sites, $t' = 7 \times 10^{-4} E_R \approx 0.01$kHz for next-nearest-neighbour hopping, and $t'' = 1.2 \times 10^{-4} E_R \approx 0.002$kHz for the hopping across neighbouring one dimensional channels of between A and B sites. The on-site Hubbard interaction is $U \approx 48 g/\lambda_L^3$ where $g/(h \lambda_L^3) = 4 \pi a_s \hbar^2 / m_a h\lambda_L^3 \approx 0.1$kHz using the $s$-wave scattering cross section $a_s$ which is 118 Bohr radii for K$^{40}$ \cite{TARRUELL2018,Esslinger2010}. The interaction strength $U$ can be independently tuned via a Feshbach resonance. Nearest-neighbour density interactions between A and B sites in the same one dimensional channel are approximately $3$Hz. Finally, the average separation between the two Wannier orbitals is on the order of $\lambda/2$. To obtain a potential difference of $V_0/(\hbar \Omega) \approx 1$, sufficient for implementing the strongly driven non-interacting effective model with $\Omega \approx 2 U/\hbar = 40 t / \hbar$, we have $V_0 = - \Omega^2 (\lambda_L/2) m_a \delta_0 \rightarrow 1 = |\Omega (\lambda_L/2) m_a \delta_0 / \hbar| \rightarrow \delta_0 \approx 2\hbar /(\Omega \lambda_L m_a) \approx 200$nm, and so requires a modulation close to the lattice spacing.

\begin{figure}[b]
	\includegraphics{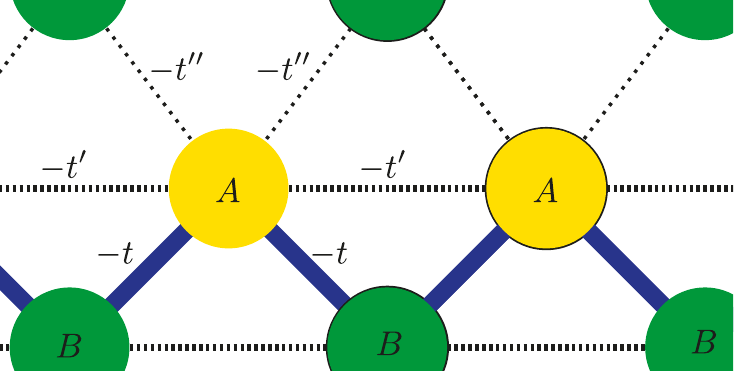}
	\caption{The dominant hopping amplitudes between sites on in a zig-zag lattice created by the potential in \fir{FIG:ZigZagPot}.}
	\label{FIG:ZigZagHop}
\end{figure}

\bibliographystyle{apsrev4-1}
\bibliography{quasicondensation.bib}

\end{document}